\pdfoutput=1
\documentclass[twocolumn, %showpacs, showkeys, linenumbers 
amsmath, amssymb, amsfonts, floatfix, superscriptaddress,notitlepage]{revtex4-1} 

\usepackage{color}
\usepackage{placeins} % floatbarrier definition
\usepackage{graphicx}
\usepackage[perpage]{footmisc}
\usepackage[normalem]{ulem}
\newcommand \Leff{\mathcal{L}_{\mathrm{eff}}}
\newcommand \Ly{L_{\mathrm{y}}}
\newcommand \Qy{\mathcal{Q}_{\mathrm{y}}}

\newcommand \keVr{\mathrm{keV_\mathrm{nr}}}
\newcommand \keVee{\mathrm{keV_\mathrm{ee}}}
\newcommand{\Xehund}{{XENON100}}

\newcommand{\Si}{\ensuremath{\mathrm{S1}}}
\newcommand{\Sii}{\ensuremath{\mathrm{S2}}}
\newcommand{\cSi}{\ensuremath{\mathrm{cS1}}}
\newcommand{\cSiib}{\ensuremath{\mathrm{cS2}_\mathrm{b}}}

\let\svthefootnote\thefootnote
\allowdisplaybreaks

\usepackage{listings}
\usepackage{color}
\usepackage{fancyvrb}

\definecolor{dkgreen}{rgb}{0,0.6,0}
\definecolor{gray}{rgb}{0.5,0.5,0.5}
\definecolor{mauve}{rgb}{0.58,0,0.82}

\newcommand{\bologna}{\affiliation{Department of Physics and Astrophysics, University of Bologna and INFN-Bologna, 40126 Bologna, Italy}}

\newcommand{\chicago}{\affiliation{Department of Physics \& Kavli Institute of Cosmological Physics, University of Chicago, Chicago, IL 60637, USA}}

\newcommand{\coimbra}{\affiliation{LIBPhys, Department of Physics, University of Coimbra, 3004-516 Coimbra, Portugal}}

\newcommand{\columbia}{\affiliation{Physics Department, Columbia University, New York, NY 10027, USA}}

\newcommand{\lngs}{\affiliation{INFN-Laboratori Nazionali del Gran Sasso and Gran Sasso Science Institute, 67100 L'Aquila, Italy}}

\newcommand{\mainz}{\affiliation{Institut f\"ur Physik \& Exzellenzcluster PRISMA, Johannes Gutenberg-Universit\"at Mainz, 55099 Mainz, Germany}}

\newcommand{\heidelberg}{\affiliation{Max-Planck-Institut f\"ur Kernphysik, 69117 Heidelberg, Germany}}

\newcommand{\munster}{\affiliation{Institut f\"ur Kernphysik, Westf\"alische Wilhelms-Universit\"at M\"unster, 48149 M\"unster, Germany}}

\newcommand{\nikhef}{\affiliation{Nikhef and the University of Amsterdam, Science Park, 1098XG Amsterdam, Netherlands}}

\newcommand{\nyuad}{\affiliation{New York University Abu Dhabi, Abu Dhabi, United Arab Emirates}}

\newcommand{\purdue}{\affiliation{Department of Physics and Astronomy, Purdue University, West Lafayette, IN 47907, USA}}

\newcommand{\rpi}{\affiliation{Department of Physics, Applied Physics and Astronomy, Rensselaer Polytechnic Institute, Troy, NY 12180, USA}}

\newcommand{\rice}{\affiliation{Department of Physics and Astronomy, Rice University, Houston, TX 77005, USA}}

\newcommand{\stockholm}{\affiliation{Oskar Klein Centre, Department of Physics, Stockholm University, AlbaNova, Stockholm SE-10691, Sweden}}

\newcommand{\subatech}{\affiliation{SUBATECH, IMT Atlantique, CNRS/IN2P3, Universit\'e de Nantes, Nantes 44307, France}}

\newcommand{\torino}{\affiliation{INFN-Torino and Osservatorio Astrofisico di Torino, 10125 Torino, Italy}}

\newcommand{\ucla}{\affiliation{Physics \& Astronomy Department, University of California, Los Angeles, CA 90095, USA}}

\newcommand{\ucsd}{\affiliation{Department of Physics, University of California, San Diego, CA 92093, USA}}

\newcommand{\wis}{\affiliation{Department of Particle Physics and Astrophysics, Weizmann Institute of Science, Rehovot 7610001, Israel}}

\newcommand{\zurich}{\affiliation{Physik-Institut, University of Zurich, 8057  Zurich, Switzerland}}

\newcommand{\paris}{\affiliation{LPNHE, Université Pierre et Marie Curie, Université Paris Diderot, CNRS/IN2P3, Paris 75252, France}}

\newcommand{\freiburg}{\affiliation{Physikalisches Institut, Universit\"at Freiburg, 79104 Freiburg, Germany}}

\newcommand\blankfootnote[1]{%
  \let\thefootnote\relax\footnotetext{#1}%
  \let\thefootnote\svthefootnote%
}

\begin{document}

\title{Effective field theory search for high-energy nuclear recoils using the \Xehund\ dark matter detector}

\author{E.~Aprile}\columbia

\author{J.~Aalbers}\nikhef

\author{F.~Agostini}\lngs\bologna

\author{M.~Alfonsi}\mainz

\author{F.~D.~Amaro}\coimbra

\author{M.~Anthony}\columbia

\author{F.~Arneodo}\nyuad

\author{P.~Barrow}\zurich

\author{L.~Baudis}\zurich

\author{B.~Bauermeister}\stockholm

% ...

\collaboration{XENON Collaboration}\email[E-mail: ]{xenon@lngs.infn.it}\noaffiliation

\author{B.~Farmer}\email[E-mail: ]{benjamin.farmer@fysik.su.se}\stockholm

\date{\today}

\begin{abstract} 

We report on WIMP search results in the \Xehund\ detector using a non-relativistic effective field theory approach. The data from science run II (34 kg $\times$ 224.6 live days) was re-analyzed, with an increased recoil energy interval compared to previous analyses, ranging from $(6.6 - 240)~\keVr$. The data is found to be compatible with the background-only hypothesis. We present 90\% confidence level exclusion limits on the coupling constants of WIMP-nucleon effective operators using a binned profile likelihood method. We also consider the case of inelastic WIMP scattering, where incident WIMPs may up-scatter to a higher mass state, and set exclusion limits on this model as well. 
\end{abstract}

\pacs{}
\keywords{Dark Matter, EFT, Xenon}

\maketitle

\section{Introduction}

Astrophysical and cosmological observations provide strong evidence that about 27\% of the energy density of the universe is made out of Dark Matter (DM). The DM hypothesis is based on the existence of a non-luminous, non-baryonic, and non-relativistic particle, the nature of which is yet unknown~\cite{Harvey1462,WMAP:9years,PLANCK}. 
Many well-motivated theoretical extensions of the Standard Model of particle physics predict the existence of one or more particles with the required properties, with masses and cross sections typically of the order of the weak scale. Such particles are collectively known as Weakly Interacting Massive Particles (WIMPs)~\cite{Bertone:2010zza}. The hypothesis that dark matter is constituted primarily of WIMPs is currently being tested by many experiments, either indirectly by seaching for evidence of their possible decay or annihilation in astrophysical processes, by searching for evidence of their direct production at collider experiments, or by directly measuring the rare scattering of astrophysical WIMPs from target nuclei in Earth-based laboratories~\cite{xe100_run_combination,PANDAX,LUXnew,COGENT,CDMSlite,CREST,DAMA}. We report on a search of this latter kind.

The traditional approach for computing predictions of the rate of WIMP-nucleon scattering has been to take only leading-order terms in a WIMP-nucleon effective field theory (EFT) with a very simple treatment of nuclear structure~\cite{LEWIN}. This leads to two main types of interactions, which are commonly labelled ``Spin Independent'' (SI) and ``Spin Dependent'' (SD). However, in recent years many authors have pointed out that in certain theories these interactions may be suppressed or nonexistent, such that otherwise subleading interactions may dominate the scattering process~\cite{Chang:2009yt}. To account for this possibility in a systematic way, a more sophisticated EFT approach has been developed ~\cite{Fitzpatrick:2012ib,Anand:MathTools,Fitzpatrick:MathTools}. In the new approach, an effective Lagrangian describing the WIMP-nucleus interaction is constructed, that takes into account all Galilean-invariant operators up to second order in the momentum exchange. This framework introduces new operators associated with different types of nuclear responses, along with the standard SI and SD ones, resulting in a set of fourteen operators $\mathcal{O}_i$ which may couple independently to protons and neutrons. In Eqs. (\ref{eq:OpDef}) we list these operators following the convention from~\cite{Anand:MathTools}. The operators depend explicitly on 4 linearly independent quantities: $\vec{v}^{\perp} \equiv \vec{v} + \frac{\vec{q}}{2\mu_N} $, the relative perpendicular velocity between the WIMP and the nucleon, $\vec{q}$, the momentum transferred in the scattering event, and $\vec{S}_\chi$, $\vec{S}_N$, the WIMP and nucleon spins. $\mathcal{O}_2$ is not considered here as it cannot be obtained from a relativistic operator at leading order.

\begingroup
\belowdisplayskip=0pt
\begin{align*}
\begin{split} 
&\mathcal{O}_1 = 1_{\chi} 1_N  \\
%&\mathcal{O}_2 = (v^{\perp})^2 \\
&\mathcal{O}_3 = i\vec{S}_N\cdot (\frac{\vec{q}}{m_N}\times\vec{v}^\perp) \\
&\mathcal{O}_4 = \vec{S}_{\chi}\cdot \vec{S}_N \\
&\mathcal{O}_5 = i\vec{S}_{\chi}\cdot (\frac{\vec{q}}{m_N}\times\vec{v}^\perp) \\
&\mathcal{O}_6 = (\vec{S}_{\chi} \cdot \frac{\vec{q}}{m_N})(\vec{S}_N \cdot \frac{\vec{q}}{m_N}) \\
&\mathcal{O}_7 = \vec{S}_N \cdot \vec{v}^\perp \\
&\mathcal{O}_8 = \vec{S}_{\chi} \cdot \vec{v}^\perp  \\
\end{split}
\begin{split}
&\mathcal{O}_9 = i\vec{S}_{\chi} \cdot(\vec{S}_N \times \frac{\vec{q}}{m_N}) \\
&\mathcal{O}_{10} = i\vec{S}_N \cdot (\frac{\vec{q}}{m_N}) \\
&\mathcal{O}_{11} = i\vec{S}_{\chi} \cdot (\frac{\vec{q}}{m_N}) \\
&\mathcal{O}_{12} = \vec{S}_\chi \cdot (\vec{S}_N \times \vec{v}^\perp) \\
&\mathcal{O}_{13} = i(\vec{S}\chi \cdot \vec{v}^\perp)(\vec{S}_N \cdot \frac{\vec{q}}{m_N})\\
&\mathcal{O}_{14} = i(\vec{S}_\chi \cdot \frac{\vec{q}}{m_N})(\vec{S}_N \cdot \vec{v}^\perp) \\
\end{split}
\end{align*}
\endgroup
\begingroup
\abovedisplayskip=0pt
\begin{align}
&\mathcal{O}_{15} = -(\vec{S}_\chi \cdot \frac{\vec{q}}{m_N})\left[(\vec{S}_N \times \vec{v}^\perp)\cdot \frac{\vec{q}}{m_N}\right]
\label{eq:OpDef}
\end{align}
\endgroup

\begin{figure}[t!]
\centerline{\includegraphics[width=1.\linewidth]{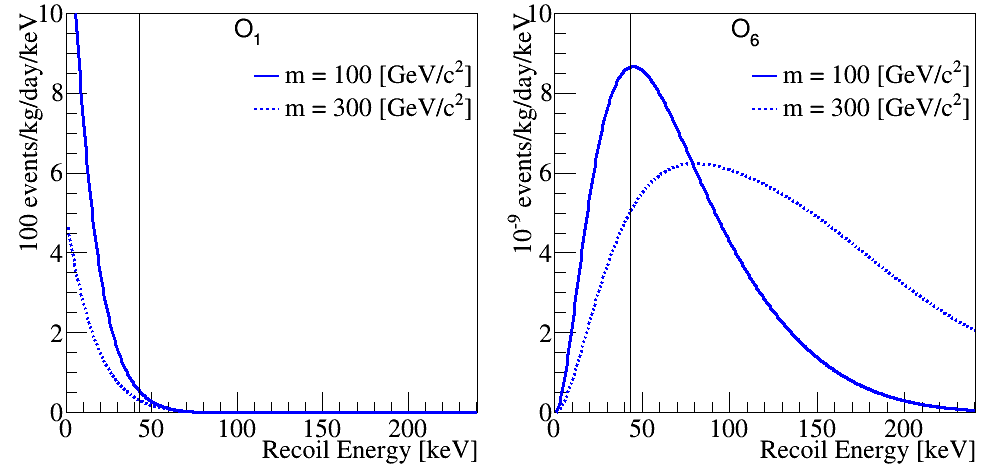}}
\caption{Example EFT recoil spectra for elastic scattering of spin-$1/2$ WIMPs on Xenon nuclei (weighted according to the isotope abundances in the XENON100 experiment). Left(right) shows the predicted spectra for EFT operator $\mathcal{O}_1$($\mathcal{O}_6$). The normalization is controlled by the coupling coefficient of each EFT operator and the experimental exposure. The solid vertical line at $43~\keVr$ shows the approximate division between the two signal regions used in this analysis. As shown, the standard SI ($\mathcal{O}_1$) spectrum is concentrated mainly in the already-explored energy region. However, some EFT operators, for certain WIMP masses, predict a significant fraction of recoil events above the upper energy cut used in the standard spin-independent analysis, motivating an extension of this cut. The highest recoil energy shown in the plots, $240~\keVr$, roughly corresponds the highest energy accounted for this analysis.}
\label{fig:dRdE}
\end{figure}

Unlike the more commonly studied types of interaction (SI,SD), which are not suppressed when $\vec{q} \rightarrow 0$ and for which the scattering rate on nucleons is expected to be largest for low energy nuclear recoils, some of the new EFT operators depend explicitly on $\vec{q}$ and so their interaction cross section is suppressed for low momentum transfers. Consequently, their scattering rate peaks at non-zero nuclear recoil energy. For sufficiently high WIMP masses, this may even occur outside typical analysis windows, which usually have an upper range of around $ 43\,\keVr$ (nuclear recoil equivalent
energy) since they are designed to search for SI and SD interactions, which predict exponentially-falling recoil spectra (see Figure~\ref{fig:dRdE}). Due to the theoretical bias of only considering SI and SD interactions, high energy nuclear recoils remain unexplored in many experiments.

	    Another typical assumption that can be relaxed is that WIMPs should scatter elastically with nuclei. There exist dark matter models in which the incoming and outgoing WIMPs have different mass states~\cite{InelasticIntro} separated by a keV-scale splitting. In the case where the outgoing state is more massive than the incoming state, the cross section for low recoil energies can again be suppressed, this time by scattering kinematics. Recently an inelastic adaptation of the EFT operator framework discussed above was developed~\cite{InelasticMath}. In this case the operators presented in Eqs.~\ref{eq:OpDef} are modified such that $\vec{v}^\perp_{inelastic} = \vec{v}^\perp_{elastic} +\frac{\delta_m}{\vert{\vec{q}}\vert^2}\vec{q}$. We consider this case in section \ref{subsubsec:Inelastic}.
	    
The EFT framework of \cite{Fitzpatrick:2012ib} is constructed at the WIMP-nucleon level and so each operator may be present independently for protons and neutrons, though UV models can of course correlate their couplings. The full EFT thus has 28 coupling parameters in addition to the WIMP mass, plus a mass splitting~$\delta$ in the inelastic case. This parameter space is too large to explore in full, so we take a similar approach to the SI/SD case 

and assume only one active operator at a time, considering it equally coupled to protons and neutrons (the ``isoscalar'' case). However, to facilitate the full exploitation of these results by the community, we provide in supplementary material a set of tools for converting any theoretical recoil spectrum $\mathrm{d}R/\mathrm{d}E$ into an accurate event rate prediction for this analysis, including all detector response and analysis efficiency effects. This may help to set a mildly conservative but quite accurate limit on arbitrary models in the full EFT parameter space, or any other particle dark matter model for which one can supply the expected recoil spectrum. These tools are described further in Appendix~\ref{app:response_table}.

Motivated by these EFT extensions of the standard WIMP framework, we report on an analysis extending the searched recoil energy range up to $240~\keVr$ for the first time in the \Xehund\ experiment, and present exclusion limits on all operators for both elastic and inelastic WIMP cases.

\section{The \Xehund\  Detector}
The \Xehund\ detector is a cylindrical dual-phase xenon (liquid and gas) time projection chamber (TPC). It is installed at the Laboratori Nazionali del Gran Sasso (LNGS) in Italy
and contains 161\,kg of liquid xenon (LXe), of which 62\,kg function as the active target ~\cite{xe100_instr2012}. 
The detector uses of a total of 178~1-inch square Hamamatsu R8520-AL photomultiplier tubes (PMTs) employed in two arrays, one in the gas phase at the top of the TPC, and the other at the bottom, immersed in the LXe. 

A particle interacting with the LXe deposits energy that creates both
prompt scintillation (\Si{}) and delayed proportional scintillation (\Sii{}) which are detected using the two PMT arrays. The \Sii{} signal is produced by ionization electrons, drifted in an electric field of $530$V/cm towards the liquid-gas interface, where they are extracted to the gas phase using a stronger electric field of $\sim12$kV/cm in which the proportional scintillation occurs. 
The spatial distribution of the \Sii{} signal on the top PMT array, together with the time difference between \Si{} and \Sii{} signals, provide respectively $x$-$y$ and $z$ position information for each interaction, allowing 3D position reconstruction to be achieved.

Interaction in different locations of  the detector have different signatures. In order to take these effects into account, a correction is applied based on light and charge collection efficiency maps. These maps are prepared using calibration sources ranging up to energies well above $240~\keVr$, which is the highest energy recoil considered in this paper. The corrected signals (\cSi{},\cSiib{}) are spatially independent and uniform to all interactions~\cite{xe100_instr2012}. Note that some of the top PMTs saturate for large \Sii{} signals and we therefore use in this analysis only the bottom PMT array to infer the energy scale in \Sii{}.

The $\Si{}/\Sii{}$ ratio is known to differ between nuclear recoil (NR) and electronic recoil (ER) interactions, and is thus used as a discriminating variable between a WIMP signal and ER background. The logarithm of this ratio, $\log(\cSiib{}/\cSi{})$ is referred later in the text as the discriminating ``$y$'' variable.

\section{Data Analysis}
\label{sec:Analysis}
In this work we re-analyze science run~II data recorded between February 2011 and March 2012, 
corresponding to 224.6~live~days. The characterization of the detector response to ER interactions is performed using dedicated calibration campaigns with $^{60}$Co and $^{232}$Th radioactive sources, while the response to NR interactions is performed using $^{241}$AmBe neutron source calibration campaigns.
 
This work extends the previous results~\cite{xe100_run10_si,xe100_run_combination}, referred to in the following as the low-energy channel, with a new study exploring the recoil energy range between $43-240~\keVr$. 
The data analysis is divided into two mutually exclusive channels, one optimized for low energies and ranging from 3-30~PE in \cSi{} (low-energy), 
the other optimized for high energies recoils ranging from 30-180~PE in \cSi{} (high-energy). These two analyses are then combined statistically. 

\subsection{Low energy channel}
\label{subsec:LowE}
This analysis channel relies on the re-analysis of run~II data described in~\cite{xe100_run_combination}. The region of interest (ROI), the background 
expectation models, data selections and their acceptances are mostly unchanged and so are only briefly summarized here. Differences with respect to said results are highlighted when present.

\begin{figure}[]
\begin{minipage}{1\linewidth}
\centerline{\includegraphics[width=1\linewidth]{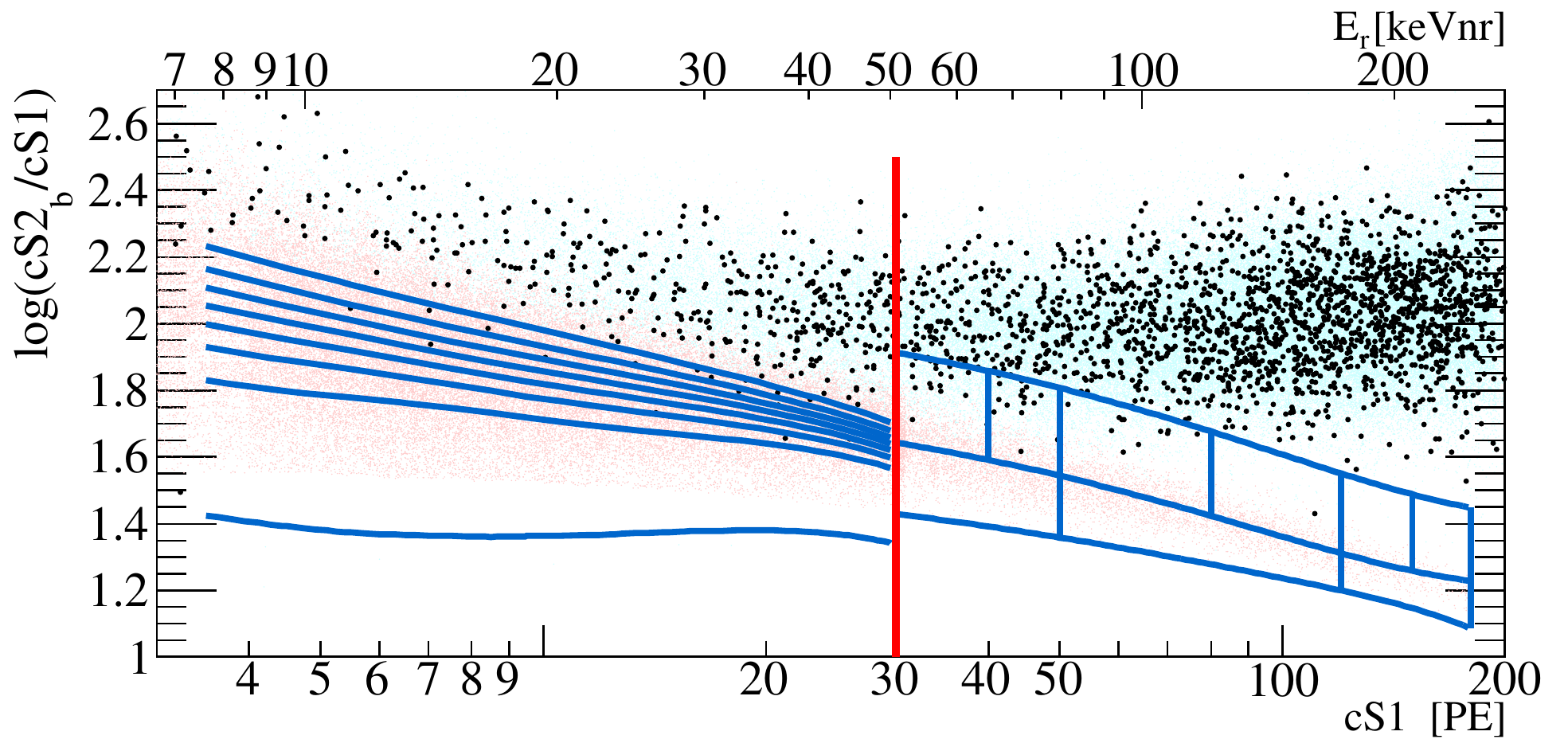}}
\end{minipage}
\caption{Summary of regions of interest, backgrounds, and observed data. ER calibration data, namely $^{60}\mathrm{Co}$ and $^{232}\mathrm{Th}$ data is shown as light cyan dots. NR calibration data ($^{241}$AmBe) is shown as light red dots. Dark matter search data is shown as black dots. The red line is the threshold between the low and high energy channels. The lines in blue are the bands. For the low-energy channel these are operator and mass dependent, but are shown here for a 50~GeV/$c^2$ WIMP using the $\mathcal{O}_1$ operator. For the high-energy region, the nine analysis bins are presented also in blue lines.
%\sout{the top left bin in this region is bin 1, the top right is bin 6, the bottom left is bin 7 and bottom right is bin 9. 
%In Sec.~\ref{sec:Results} we show similar data but for regions above the upper range of this analysis, going up to 1000\,PE in cS1, for completeness as part of final unblinding of XENON100 data.} 
}
\label{fig:phasespace}
\end{figure}

The ROI for this channel is defined in the ($y,\cSi{}$)-plane and is shown in Figure~\ref{fig:phasespace}.  The lower 
bound on $y$ corresponds to a 3\,$\sigma$ acceptance quantile (as a function of \cSi{}) of a 20~GeV WIMP mass signal model assuming an $\mathcal{O}_1$ (SI) interaction, while the upper bound is fixed at $y=2.7$.
The range in \cSi{} is selected as 3 to 30\,PE. 
The ROI is further divided into eight sub-regions (also called bands) depending on the operator $\mathcal{O}_i$ and on the WIMP mass hypothesis. 
These bands are arranged to achieve constant expected signal density in each region, as described in~\cite{xe100_run_combination}.

Other than falling into the ROI, an event should fulfill several additional selection criteria (cuts). Data quality and selection cuts are defined to remove events with poor data quality or noisy signals. Events are discarded if they present a time-coincident signal in the outer LXe veto, \Sii{} signals below threshold, multiple-scatters, or are localized outside a predefined fiducial volume of 34 kg. In addition, this analysis channel uses the post-unblinding cuts and data reprocessing described in~\cite{xe100_run_combination}. More details on these selection criteria and their relative WIMP signals acceptances can be found in~\cite{Aprile:2012vw,xe100_run_combination}. 

%%%%%%%%%%%%%%%%%%%%%%%%%%%%%%%%%%%%%%%%%%%%%
Note that this analysis channel does not employ a variable lower \Si{} threshold as a function of the event position in the TPC, but instead applies a fixed lower threshold cut on \cSi{} at 3\,PE, conversely to the choice made in~\cite{xe100_run_combination}.

The expected background is modeled separately for ER and NR contributions which are then scaled to exposure and added together.
The NR background is estimated by Monte Carlo simulation and accounts for the radiogenic and cosmogenic neutron
contributions~\cite{Aprile:2013tov}. The ER background is parametrized as the linear combination of Gaussian-shaped and non-Gaussian components.
The former is obtained via a parametric fit of the $^{60}$Co and $^{232}$Th calibration data, as discussed in~\cite{xe100_run10_si}.

The latter, which consist of anomalous events such as those 
presenting incomplete charge collection or accidental coincidence of uncorrelated \Si{}s and \Sii{}s,  
is evaluated via dedicated techniques described in~\cite{xe100_run_combination}.

Systematic uncertainties on the background model arising from the Gaussian parametrized fit, and from the normalisations of the NR and non-Gaussian components, have been evaluated and propagated to each band. 
These errors are small with respect to the statistical uncertainties of each band, which are conservatively taken as the overall uncertainty~\cite{xe100_run_combination}, as discussed in Sec.~\ref{sec:LikelihoodFunction}.

\subsection{High energy channel}
\label{subsubsec:HighE}
This analysis channel targets high energy nuclear recoils and is the focus of this work. The data selection criteria used are based on the criteria described in detail in \cite{Aprile:2012vw}, which were optimized for high acceptance to low energy nuclear recoils. Most of these cuts were found to be fully compatible with (or easily extended) to high energy depositions, however some required more comprehensive studies, which are described in the following . 

The width of an \Sii{} pulse increases with the depth (z) of the interaction. This is due to the diffusion of the electron cloud during its propagation
through the liquid xenon. Since low energy \Sii{} events show larger spread
due to low statistics of drifted electrons, the cut was previously defined in an energy-dependent way. However, for the large recoil energies considered in this channel, this energy dependency is no longer valid. We therefore use here a cut on the \Sii{} width which is a function of the depth of the interaction alone. 

As a WIMP will interact only once in the detector, we remove events which have more than one \Sii{}. We adopt in this analysis a cut that is more suitable to higher energies and demand a single \Sii{} in a 160 $\mu$s window, instead of a linear dependence between the second \Sii{} size and the first. 

To define the interaction's exact location in ($x,y$), we use several algorithms, one of which is based on a Neural Network (NN)~\cite{Aprile:2012vw}. The NN was not trained to recognize high energy ER events and therefore a cut on the NN reconstruction quality is not suitable for this analysis. We therefore discard this cut but keep all other selections on position reconstruction quality, which is sufficient to ensure a correct position reconstruction. 

\begin{figure}[t!]
\begin{minipage}{0.9\linewidth}
\centerline{\includegraphics[width=1.\linewidth]{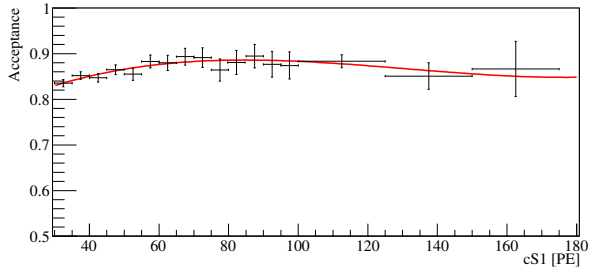}}
\end{minipage}
\caption{The total acceptance of all cuts used. Data from calibration is shown in black, with a 3rd order polynomial fit in red.}
\label{fig:Acc}
\end{figure}

The total acceptance to WIMP signals is computed based on $^{241}$AmBe calibration data as a function of \cSi, following the procedure described in~\cite{Aprile:2012vw}. We present this function in Figure~\ref{fig:Acc}, where the total acceptance is fitted using a third order polynomial.

We define our signal region in the discrimination $(y,\cSi)$-plane using $^{241}$AmBe calibration data. 
The region of interest is shown in Figure~\ref{fig:phasespace} as blue contour lines. The upper bound in $y$ is defined such that the contribution due to xenon inelastic interaction lines is negligible. The lower bound is defined as the 3\,$\sigma$ acceptance quantile of the $^{241}$AmBe distribution.

We divide our signal region into two bands in $y$, constructed such that the $^{241}$AmBe data sample is equally distributed in between them. The number of events in each band is $\sim3000$. The bands are further divided into nine bins, the number and boundaries of which have been optimized via Monte-Carlo (MC) simulation. The definitions of the bins boundaries are presented in Table~\ref{table:BinDef} and in Figure~\ref{fig:phasespace}. 

The main source of background results from ER leakage. We therefore estimate the background distribution in the ROI using $^{60}$Co and $^{232}$Th calibration events.  
Contributions from radiogenic and cosmogenic neutrons, as well as accidental coincidence, are negligible for such a high energy recoil. In Table~\ref{table:BinDef} we report the 
background expectation in the ROI along with the observed events for each bin.
Here the background expectation is computed by scaling the calibration sample yield by $6.54\times10^{-3}$, which is the ratio of observed counts to calibration counts in an independent sideband. The sideband is defined above the upper limit of this analysis and below the ER calibration band mean. Note that in the computation of exclusion limits, the background normalization is fitted to data, rather than using the sideband normalization, as described in section~\ref{sec:LikelihoodFunction}. 

%%%%%%%%%%%%%%%%%%%%%%%%%%%%%%%%%%%%%%%%%%%%%%%%%%%%%%%%%%%%%%%%%%%%%%%%%%%%%%%%%%%%%%%%%%%%%%%%%%%%%%%

\begin{table}
\resizebox{1.\columnwidth}{!}{

	 \begin{tabular}{ c c c c c c } 
 \hline\hline
 \# & Band  & Energy Range $(\cSi)$  & \# Background Events & \# Data Events \\  
 \hline
 1 & upper & 30  - 40  & 24$\pm$5 & 20 \\ 
 
 2 & upper & 40  - 50  & 16$\pm$3 & 17 \\
 
 3 & upper & 50  - 80  & 12$\pm$3 & 11 \\
 
 4 & upper & 80  - 120 & $1.1\pm0.3$  & 1  \\
 
 5 & upper & 120 - 150 & $(1.0\pm0.5)\times 10^{-1}$  & 1  \\  
 
 6 & upper & 150 - 180 & $(0.8\pm0.4)\times 10^{-1}$ & 0  \\  
 
 7 & lower & 30  - 50  & $0.9\pm0.3$  & 0  \\  
 
 8 & lower & 50  - 90 & $(3.5\pm1.2)\times 10^{-1}$ & 0  \\  
 
 9 & lower & 120 - 180 & $(1.8\pm0.7)\times 10^{-1}$& 0  \\  
 \hline\hline
\end{tabular}
}

\caption{Definitions and contents of the analysis bins for the high energy channel. The expected background counts are calculated by taking the calibration sample and scaling it by $6.54\times10^{-3}$, which is the ratio of observed counts to calibration counts in a sideband.}  \label{table:BinDef} 
\end{table}

\subsection{Signal model}
\label{subsec:SignalModel}
The signal model is produced by taking a theoretical event rate spectrum, the production of which is described in sections \ref{subsubsec:Elastic} and \ref{subsubsec:Inelastic}, and applying the analysis acceptance and detector response as described in ~\cite{Aprile:2012vw}  to obtain the expected event rate in the detector in terms of detector variables (i.e. \cSi{}, \cSiib{}). 
In both analysis channels, we use Eq.~\ref{eq:LeffEnergyScale} in order to compute the expected average \cSi{} for a given NR energy,
\begin{equation}
\label{eq:LeffEnergyScale}
	\langle \cSi \rangle = E_{\mathrm{nr}} \cdot (\Ly \Leff) \cdot   \left(\frac{S_\mathrm{nr}}{S_\mathrm{ee}}\right) 
\end{equation}

where $E_\mathrm{nr}$ is the recoil energy, $\Ly$ is the average light yield in the detector, $\Leff$ is the scintillation efficiency relative to 122$\keVee$ as a function of $E_\mathrm{nr}$, and $S_\mathrm{ee}$ and $S_\mathrm{nr}$ are the quenching factors due to the externally applied electric field. Aside from $E_\mathrm{nr}$ and $\Leff$ these parameters have fixed values, namely $\Ly = 2.28 \pm 0.04$, $S_\mathrm{nr} = 0.95$, and $S_\mathrm{ee} = 0.58$. Recoils below $3~\keVr$ are assumed to produce no light. For details of the physics behind these parameters and the construction of the signal probability density function (PDF) please see \cite{Aprile:2012vw,xe100_run_combination}. 

For the low-energy region, the expected \cSiib{} signal is computed following~\cite{DataMCXenon} using Eq.~\ref{eq:Qy},

\begin{equation}
\label{eq:Qy}
	\langle \cSiib \rangle = E_{\mathrm{nr}}\Qy Y   
\end{equation}
where $Y = 8.3 \pm 0.3$ 
is the amplification factor determined from the detector response to single electrons~\cite{XenonSingleElectron}, and $\Qy$ is the charge yield as a function of $E_\mathrm{nr}$. Applying the detector and PMT responses, and the acceptance as in \cite{xe100_run_combination}, defines the low-energy signal model over the region $3~\mathrm{PE} < \cSi{} < 30~\mathrm{PE}$, with $\cSiib{} > 73.5~\mathrm{PE}$ as the \Sii{} threshold.

Eq. \ref{eq:Qy} hides a subtlety. The actual \cSiib{} PDF is composed of two pieces, a Poisson term associated with the initial charge liberation and a Gaussian term associated with the PMT response and other detector effects:
\begin{equation}
\label{eq.cS2pdf}
p_\mathrm{S2}(\mathrm{\cSiib}|E) = \sum_{N'} P_\mathrm{pmt}(\mathrm{\cSiib}|Y N',\sigma_Y \sqrt{N'})\cdot\mathrm{Pois}(N'|\mu_Q)
\end{equation}
where $\mu_Q=E_{\mathrm{nr}}\Qy$ is the expected number of liberated charges in a nuclear recoil event of energy $E$, and $N'$ is the actual number of liberated charges. The amplification factor $Y$ is applied to the actual number of liberated charges $N'$, not the expected number $\mu_Q$. Associated with this is the variance of the Gaussian response PDF, $\sigma_Y\sqrt{N'}$, where in this analysis $\sigma_Y = 6.93$ as measured and described in~\cite{XenonSingleElectron}. 
%%%%%%%%%%%%%%%%%%%%%%%%%%%%%%%%%%%%%%%%%%%%%%%%%%%%%%%%%%%%%%%%%%%%%%%%%%%%%%%%%%%%%%%%%%%%%%%%%%%%%%%%%%%%%%%%%%

For the high energy region we cannot produce the \Sii{} distribution in the same way as the method in~\cite{DataMCXenon}, since it  has not been calibrated for such high recoil energies. We therefore use the NR calibration data distribution in log($\mathrm{\cSiib/\cSi}$) to estimate the WIMP distribution. Above 180~PE in \cSi{}, the event yield of $^{241}$AmBe data is too low to estimate the distribution accurately. This forms the upper bound of this analysis. With the \cSiib{} distribution determined by this empirical method, we require only a prediction of the \cSi{} distribution. This is obtained from Equation (\ref{eq:LeffEnergyScale}), followed by the application of detector and PMT responses, as well as the acceptance given in Figure~\ref{fig:Acc}, which completes the high-energy signal model definition.

\begin{figure}[h!]
\begin{minipage}{1.\linewidth}
\centerline{\includegraphics[width=1.\linewidth]{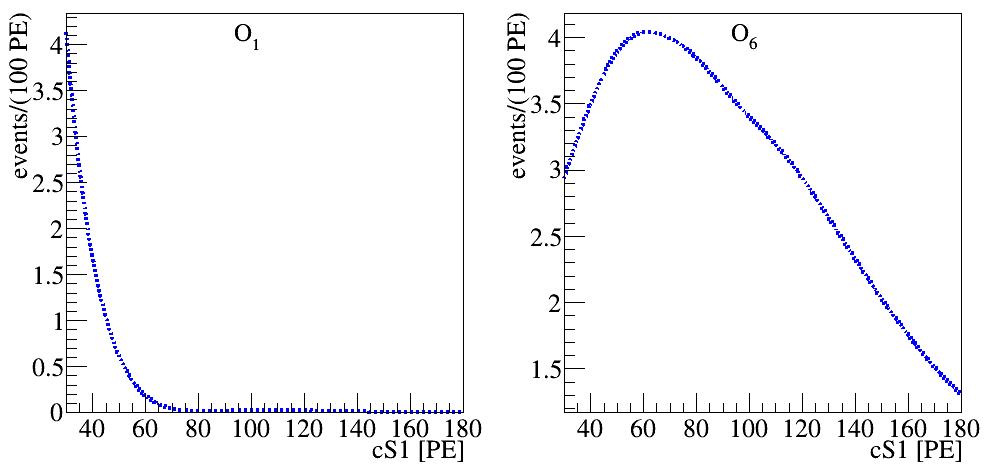}}
\end{minipage}
\caption{The expected signal in the high energy region for a 300~GeV/$c^2$ WIMP mass, normalized to 5 events. Left(right) is the spectra for $O_1$($O_6$). Notice that for $O_1$ most of the events are not expected to deposit energy higher than 30~PE whereas for $O_6$ a large fraction of the events appear in this region.}
\label{fig:HighE}
\end{figure}

\begin{figure}[h!]
\begin{minipage}{1.\linewidth}
\centerline{\includegraphics[width=1.\linewidth]{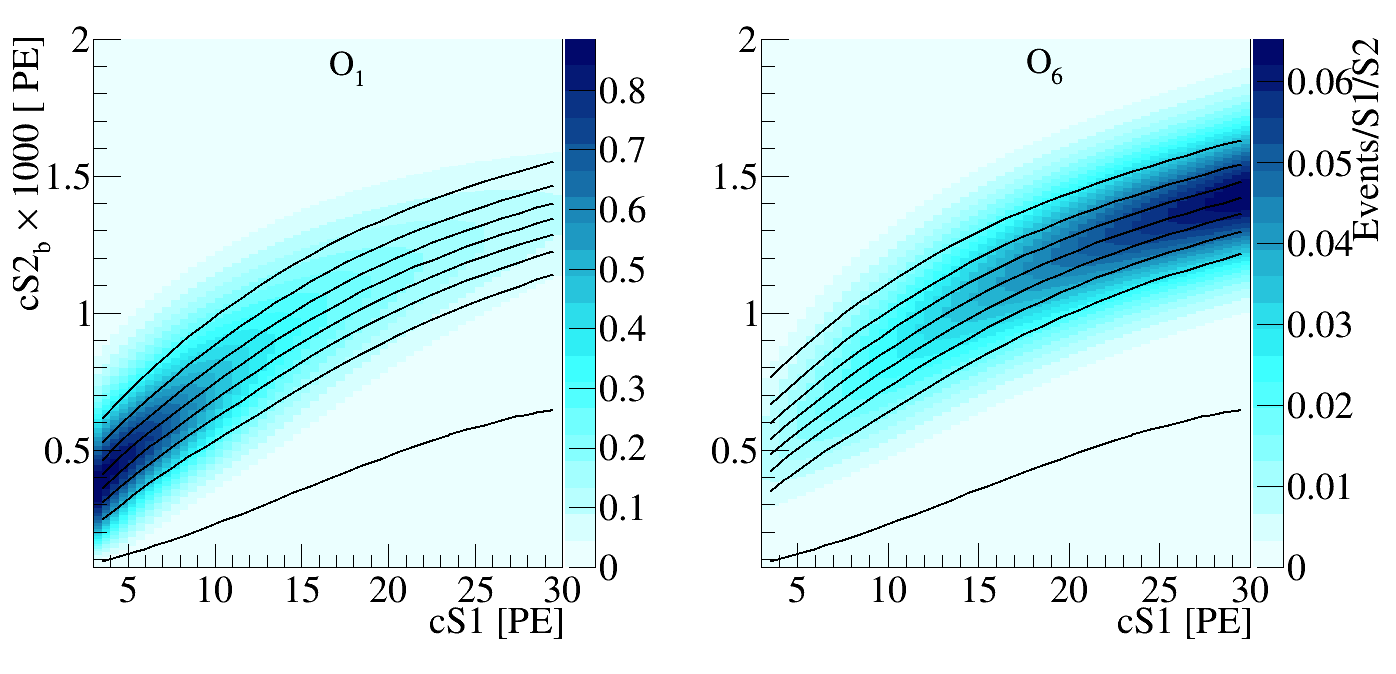}}
\end{minipage}
\caption{The expected signal in the low energy region for a 300~GeV/$c^2$ WIMP mass, normalized to 5 events. Left(right) is the spectra for $\mathcal{O}_1$($\mathcal{O}_6$). Notice that for $\mathcal{O}_1$ most of the events are expected to deposit energy lower than 30~PE whereas for $\mathcal{O}_6$ a large fraction of the events do not appear in this region at all. The black lines indicate the bands constructed on these specific mass and operator models, and are dividing the signal into 8 equally distributed signal sub-regions. This parameter space can be mapped with a one to one mapping to the $(y-\cSi)$ space.}
\label{fig:LowE}
\end{figure}

Figures ~\ref{fig:HighE} and \ref{fig:LowE} shows dashed signal distribution examples for two EFT operators and for the low and the high energy region, respectively.
In both cases, the signal distributions are normalized to yield 5 events in the total energy range (low-energy and high-energy).

\subsubsection{Elastic scattering}
\label{subsubsec:Elastic}

The expected recoil energy spectrum of each WIMP mass for each EFT operator is calculated using the Mathematica package \texttt{DMFormFactor} supplied by Anand et. al.~\cite{Fitzpatrick:MathTools,Anand:MathTools}. We use standard assumptions as in previous analyses (e.g \cite{xe100_run_combination}) regarding the local dark matter density and velocity distribution, namely $\rho_\mathrm{local} = 0.3$~GeV$\cdot c^{-2}$/$\mathrm{cm}^{3}$ and a Maxwell-Boltzman distribution with a mean given by the local circular velocity $v_0 = 220$ km/s and cut off at an escape velocity of $v_\mathrm{esc} = 544$ km/s. The responses of xenon nuclei to a scattering event are computed from one-body density matrices provided with the package, in contrast to the Helm form factors which have been used in previous analyses. These spectra are produced for the seven most abundant xenon isotopes (128, 129, 130, 131, 132, 134 and 136), combined in proportion to the abundance of these isotopes in the XENON detector \cite{xe100_run10_sd}, then translated into expected signal rates via the method described above.

\subsubsection{Inelastic WIMP scattering}
\label{subsubsec:Inelastic}
To obtain recoil spectra for WIMP-nucleon scattering for all EFT operators with inelastic kinematics, we use a modified version of \texttt{DMFormFactor} provided by Barello et. al. \cite{InelasticMath}. The authors have modified the original package to enforce the new energy conservation condition $\delta_m + \vec{v}\cdot\vec{q} + \left|\vec{q}\right|^2/2\mu_N = 0$, primarily by replacing 
$\vec{v}^\perp_{elastic} \rightarrow \vec{v}^\perp_{inelastic} = \vec{v}^\perp_{elastic} +\frac{\delta_m}{\vert{\vec{q}}\vert^2}\vec{q}$ in the definitions of the EFT and nuclear operators, giving rise to the well-known minimum velocity for scattering
\begin{equation}
  v_\mathrm{min}/c = \frac{1}{\sqrt{2 m_N E_R}} \left|\frac{m_N E_R}{\mu_N} + \delta_m\right|
\end{equation}
where $\mu_N$ is the WIMP-nucleon reduced mass.

Assumptions regarding the dark matter halo and nuclear physics are unchanged. The mass splitting $\delta_m$ between dark matter states is varied from 0 to 300 keV, safely beyond the value at which the predicted rate is zero for the entire mass range we consider.

\subsection{Statistical inference}
\label{sec:LikelihoodFunction}
The statistical interpretation of data is performed using a binned profile likelihood method, in which hypothesis testing relies upon a likelihood ratio test statistic, $\tilde{q}$, 
and its asymptotic distributions~\cite{asympt}. The two analysis channels are combined by multiplying their likelihoods together to produce a joint likelihood. 
Both analyses parametrize the NR relative scintillation efficiency, $\Leff$, based on existing measurements~\cite{run8Result}. Its uncertainty is the major contributor to energy scale uncertainties and is considered as correlated between the two analysis channels via a joint nuisance likelihood term.
Throughout this study, all the parameters related to systematic uncertainties are assumed to be normally distributed.

For the low energy channel an extended likelihood function is employed which is very similar to the one reported in~\cite{Aprile:2011hx} and described in detail in~\cite{xe100_run_combination}. 
The $(y,\cSi{})$-plane is divided into eight WIMP mass dependent bands where events are counted. This binned approach is extended with the corresponding \cSi{}-projected PDF of each band. The total normalization of the background is fit to data, and an uncertainty is assigned to the relative normalization of each band according to the corresponding statistical uncertainty of the calibration sample.
Signal shape variations due to energy scale uncertainty are modeled via simulation. These include  the said $\Leff$ uncertainties and additionally 
the charge yield uncertainties, which are parametrized based on $\Qy$ measurement as described in~\cite{DataMCXenon}.

The high energy channel analysis employs a binned likelihood function. Observed and expected event yield are compared in the nine ROI $(y,\cSi{})$-bins described in section~\ref{subsubsec:HighE}. 
Given the large statistical uncertainty of the background model the above extended likelihood approach is not repeated here.
Instead, the maximum likelihood estimation of the background expectation in each bin is constrained by the statistical uncertainty of the calibration sample, while the total 
normalization is fit to the data. Additionally, to account for potential mismodeling of the expected background distribution, mainly due to anomalous multiple scatter events,
a systematic uncertainty of 20\% is assigned independently to each bin. In the high energy channel, uncertainty on the signal acceptance of analysis selections are computed for each signal hypothesis using the parametrized acceptance curve shown in Figure~\ref{fig:Acc}.
Uncertainties on the signal model $(y,\cSi{})$ distribution due to $^{241}$AmBe sample statistical fluctuations, as well as energy scale shape variation due to $\Leff$ uncertainties, are taken into account.

\section{Results}
\label{sec:Results}
%\subsection{Elastic Scattering}

A benchmark region of interest is defined between the upper and lower thresholds in \cSi{} for each channel. This region
is bounded in $y$-space from above by the $^{241}$AmBe NR mean line and below by the lower 3$\sigma$ quantile of the $^{241}$AmBe neutron calibration data. The expected background in the region is $3.0 \pm 0.5_{stat}$ (low-energy) and $1.4 \pm 0.3_{stat}$ (high-energy). The number of DM candidates in this benchmark region is 3 (low-energy), and 0 (high-energy). Consequently, the data is compatible with the background-only hypothesis and no excess is found. 

For the elastic scattering case, a 90\%\,CL$_S$~\cite{cls} confidence level limit is set on the effective coupling constant, $c_i$,  for all operators and masses in the range of 10~GeV/$c^2$ to 1 TeV/$c^2$. The $c_i$ are dimensionful, with units of $[\mathrm{mass}]^{-2}$, so we first convert them to dimensionless quantities by multiplying them by $m_\mathrm{weak}^2=(246.2\text{ GeV})^2$, following the conventions of \cite{Anand:MathTools}. 

\begin{figure*}
\begin{minipage}{1.\linewidth}{}
\centerline{\includegraphics[width=\textwidth,height=0.99\textheight,keepaspectratio]{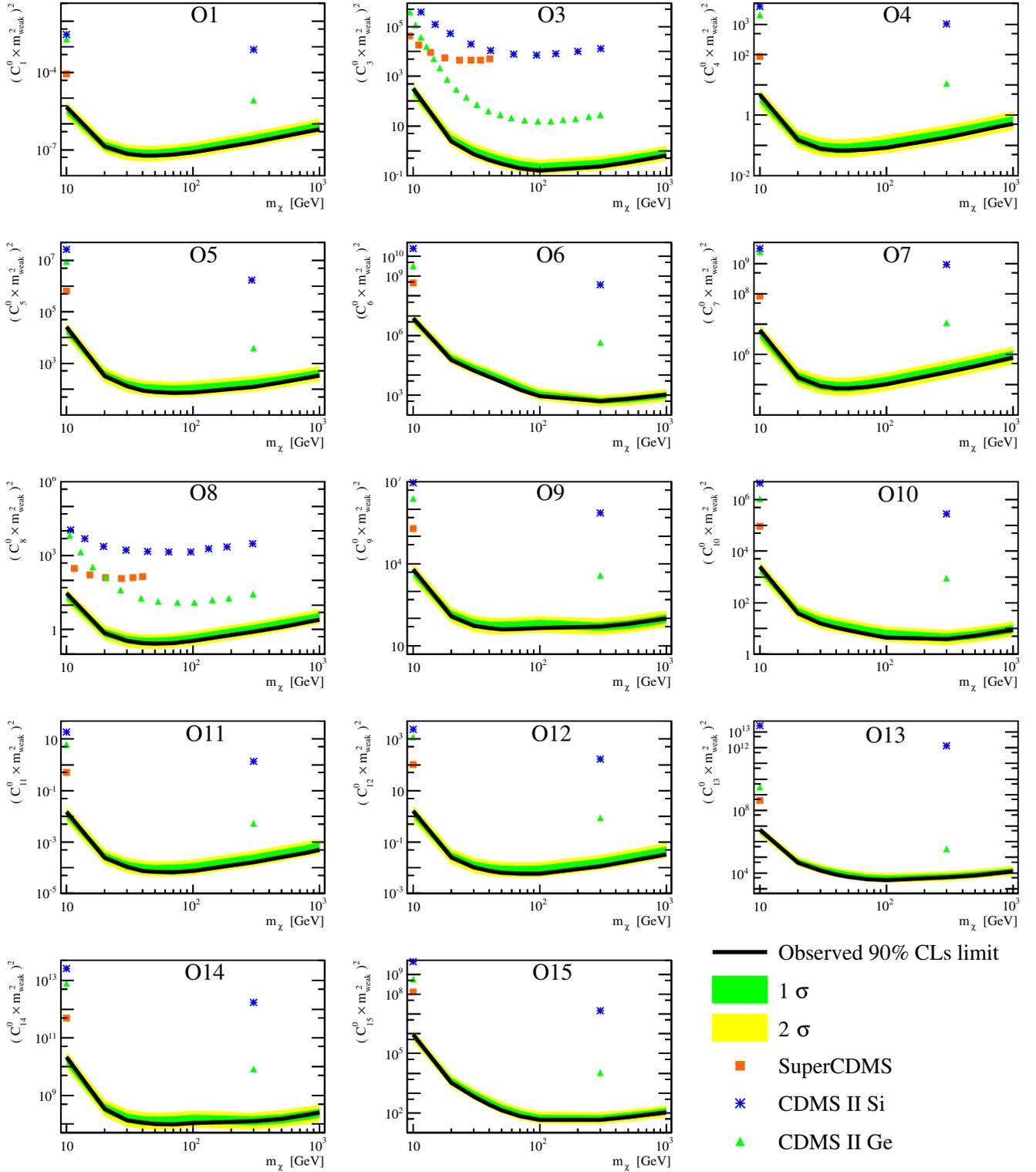}}
\end{minipage}
\caption{The \Xehund\ limits (90\%\,CL$_S$) on isoscalar dimensionless coupling for all elastic scattering EFT operators. The limits are indicated in solid black. The expected sensitivity is shown in green and yellow(1$\sigma$ and 2$\sigma$ respectively). Limits from CDMS-II Si, CDMS-II Ge, and SuperCDMS \cite{CDMSEFT} are presented as blue asterisks, green triangles, and orange rectangles, respectively (color online). For operator 3 and 8 a full limit was published, for all other operators only $m_\chi = 10$ and $m_\chi =300$ are available.}
\label{fig:elasticLimit}
\end{figure*}

These limits are shown in Fig.~\ref{fig:elasticLimit} in black, along with limits from CDMS-II Si, CDMS-II Ge and SuperCDMS~\cite{CDMSEFT}.

\begin{figure}
\centerline{\includegraphics[width=1.\linewidth]{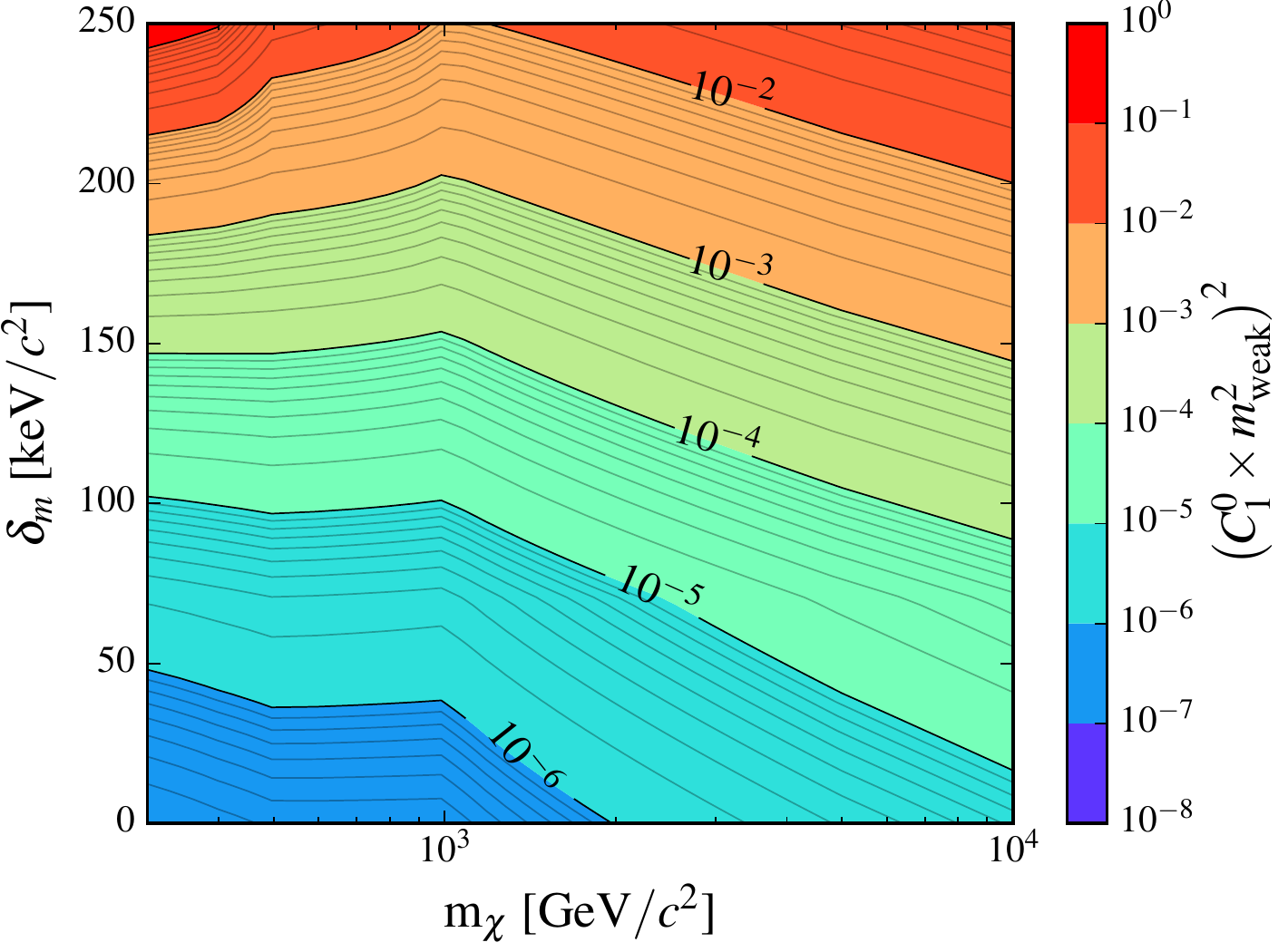}}
\caption{90\%\,CL$_S$ limits, for the inelastic model, on the magnitude of the coupling constant for $\mathcal{O}_1$, reported as a function of the WIMP mass and mass splitting $\delta$.}
\label{fig:O1Inel}
\end{figure}

\begin{figure*}
\begin{minipage}{1.\linewidth}
\centerline{\includegraphics[width=\textwidth,height=0.99\textheight,keepaspectratio]{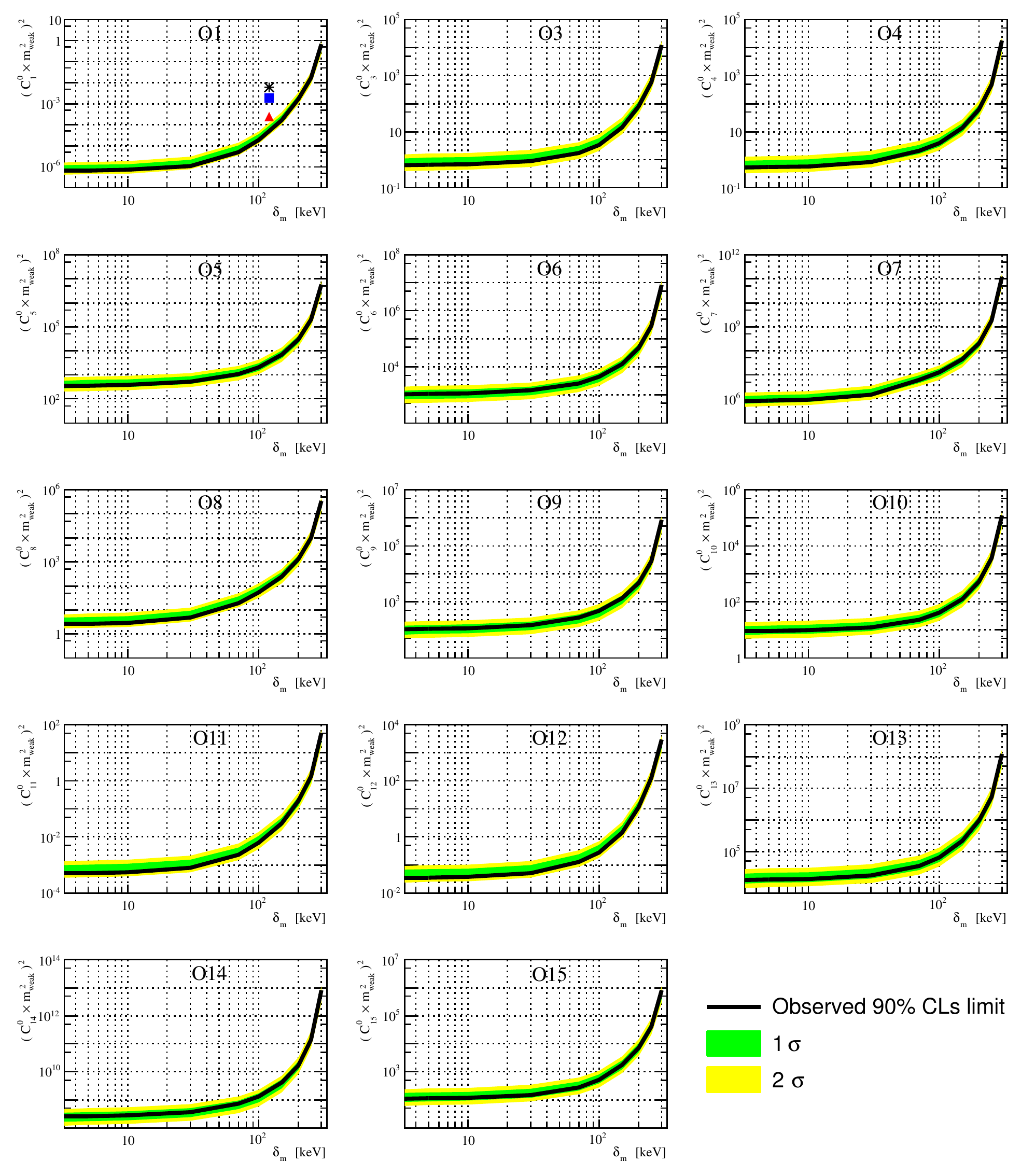}}
\end{minipage}
\caption{The \Xehund\ 90\%\,CL$_S$ limits on a 1 TeV/$c^2$ WIMP isoscalar dimensionless coupling constant as function of the WIMP mass splitting $\delta_m$  for all inelastic scattering EFT operators. Limits are indicated in solid black. The expected sensitivity is shown in green and yellow (1$\sigma$ and 2$\sigma$ respectively). For $\mathcal{O}_1$ (SI) results from \Xehund(red triangle) CDMS-II(blue rectangle) and ZEPLIN-III(black star) are overlaid.}
\label{fig:InelasticLimit}
\end{figure*}

For the inelastic scattering case,  90\%\,CL$_S$ confidence level limits on the coupling constants 
(again scaled by $m_\mathrm{weak}^2$) are set. Fig.~\ref{fig:O1Inel} shows limits on the $\mathcal{O}_1$ (SI) coupling constant as a function of mass splitting and WIMP mass, Fig.~\ref{fig:InelasticLimit} shows limits for all other operators as a function of the mass splitting $\delta_m$ with a fixed WIMP mass of 1 TeV/$c^2$,  
projections of results from CDMS-II~\cite{CDMS_Inelastic}, ZEPLIN-III~\cite{Zepplin_Inel}, and \Xehund~\cite{XENON_Inelastic_WIMP} in the coupling constant and $\delta_m$ parameter space are also reported.

For the elastic operator $O_1$ our results can be compared to those of standard SI analyses by computing the relevant zero-momentum WIMP-nucleon cross-sections. This is not simple to do rigorously because the treatment of nuclear structure used in our analysis is different than in standard analyses, however this difference is small for scattering via $O_1$. We can therefore quite safely use the `traditional' correspondence~\cite{DeSimone:2016fbz}
\begin{equation}
\sigma_{N}^\mathrm{SI} = \left(C^N_1\right)^2 \frac{\mu_{\chi,N}^2}{\pi}
\end{equation}
where $\mu_{\chi,N}$ is the WIMP-nucleon reduced mass. Standard SI analyses assume isospin-conserving interactions, as we do in this analysis, so we can simply set $C^N_1 = C^0_1$, such that $\sigma_{p}^\mathrm{SI}=\sigma_{n}^\mathrm{SI}$. 

In principle a similar comparison can be done between our limit on the $O_4$ coupling and standard SD analysis limits, however this time the standard analyses do {\em not} assume isospin-conserving interactions. Instead they typically assume maximal isospin violation, that is, assuming that WIMPs couple either protons or neutrons. Limits are then derived independently on $\sigma_{p}^\mathrm{SD}$ and $\sigma_{n}^\mathrm{SD}$. Because of this difference in assumptions, our limits on SD couplings are not directly comparable to usual analyses. However, they can be recast under the appropriate alternate model assumptions using the detector response tables we provide in the supplementary material.

\section{Summary}
We have shown the first analysis of \Xehund\ data at recoil energies above 43 keV, with the new high energy bound set to 240 keV. We considered in this paper two models which predict interactions in this energy region: an EFT approach for elastic WIMP-nucleon scattering, and a similar EFT approach but considering instead inelastic WIMP-nucleon scattering. The observed data was compatible with background expectations, and 90\%\,CL$_S$ exclusion limits were constructed for WIMP masses between 10-1000 GeV.

\begin{acknowledgments}
We would like to thank Andrew Liam Fitzpatrick and Spencer Chang for supplying and helping with their Mathematica packages . We gratefully acknowledge support from the National Science Foundation, Swiss National Science Foundation, Deutsche Forschungsgemeinschaft, Max Planck Gesellschaft, German Ministry for Education and Research, Netherlands Organisation for Scientific Research, Weizmann Institute of Science, I-CORE, Initial Training Network Invisibles (Marie Curie Actions, PITNGA-2011-289442), Fundacao para a Ciencia e a Tecnologia, Region des Pays de la Loire, Knut and Alice Wallenberg Foundation, Kavli Foundation, and Istituto Nazionale di Fisica Nucleare. We are grateful to Laboratori Nazionali del Gran Sasso for hosting and supporting the XENON project.
\end{acknowledgments}

%\newpage

\appendix*

\section{SIGNAL MODEL DETECTOR RESPONSE TABLE}
\label{app:response_table}

\begin{figure}
\centerline{\includegraphics[width=1.\linewidth]{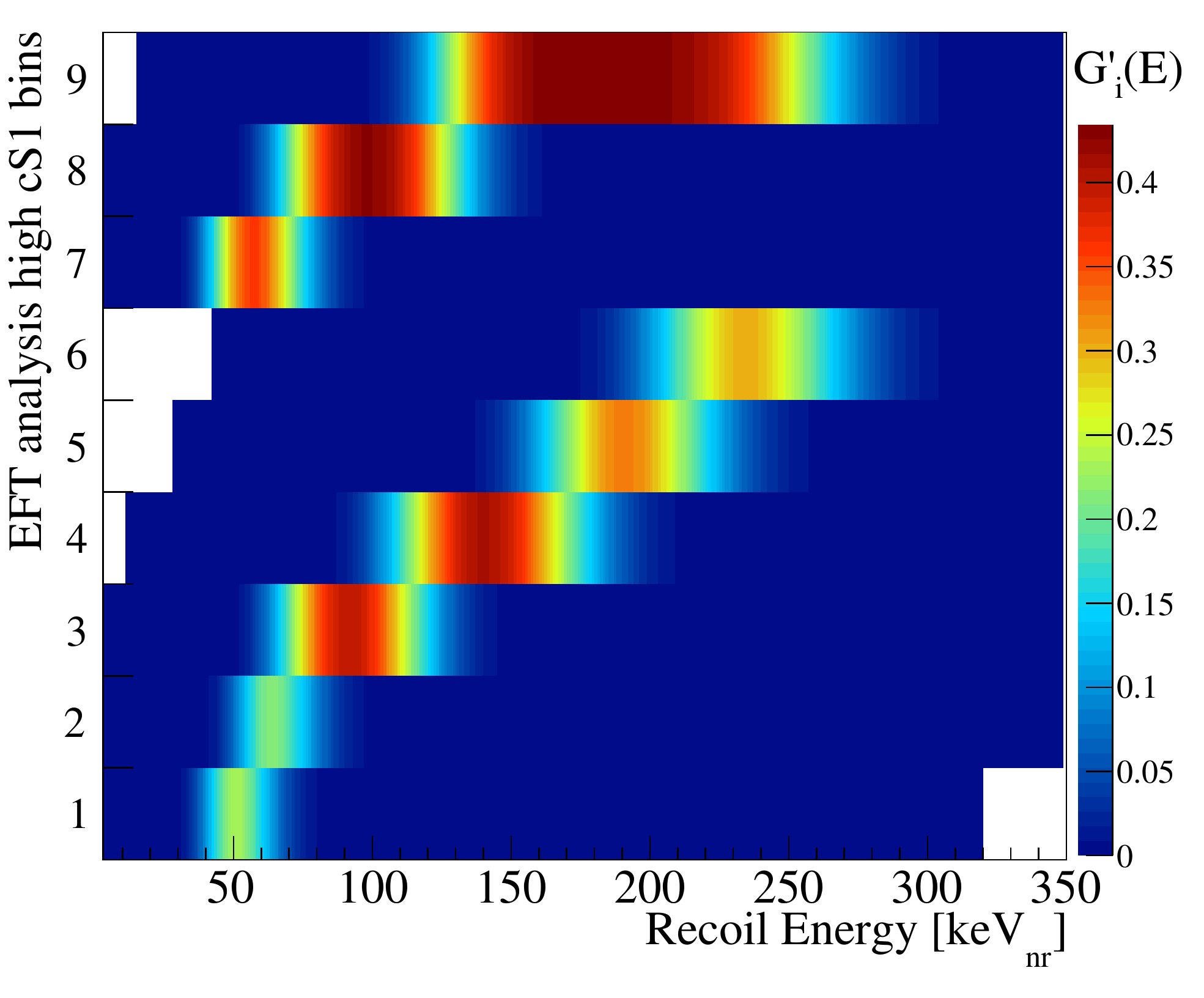}}
\caption{A visualization of the detector response table for $-1\sigma$ (i.e. conservative) $\Leff$, as provided in the supplementary material. The y axis indicates the bins used for the high-energy signal region of this analysis (explained in ~\ref{table:BinDef}). The $x$ axis shows recoil energies, and the colors give the probability density for a recoil of a given recoil energy to produce an event in each analysis bin. To produce a signal model for this analysis, one simply multiplies the table values by $\mathrm{d}R/\mathrm{d}E$ and integrates over $E$. The result is the predicted signal rate for each analysis bin.}
\label{fig:smeartable_highE}
\end{figure}

In this appendix we describe digital tables which can be used to construct an accurate signal model for this analysis given any input recoil spectrum $\mathrm{d}R/\mathrm{d}E$ arising from a theoretical model. A visualization of the tables is shown in Fig.~\ref{fig:smeartable_highE}, and in section \ref{app:example_code} we show a simple example Python code of how to use the supplied tables. Currently we provide these tables only for the high-energy analysis region.

The signal model for the high-energy analysis region can be expressed analytically in the form:
\begin{align}
\label{eq:high2D}
  \frac{\mathrm{d} R}{\mathrm{d}\cSi} &= \int \! \frac{\mathrm{d}R}{\mathrm{d}E} \cdot \epsilon_\mathrm{S1}(\cSi) \cdot \epsilon_\mathrm{S2'}(E) \cdot p_\mathrm{S1}(\mathrm{\cSi}|E) \, \mathrm{d}E \\
  &= \int \! \frac{\mathrm{d}R}{\mathrm{d}E} G(\cSi,E) \, \mathrm{d}E
\end{align}
where $\epsilon_\mathrm{S1}(\cSi)$ and $\epsilon_\mathrm{S2'}(E)$ represent analysis cut efficiencies, $p_\mathrm{S1}(\mathrm{\cSi}|E)$ encodes detector effects, and $\mathrm{d}R/\mathrm{d}E$ gives the theoretically predicted nuclear recoil rate from WIMP scattering. In the second line we emphasis that all the detector and analysis effects can be encoded in a single function $G(\cSi,E)$. To make a signal prediction for the bins in our analysis, this expression needs to be integrated over the appropriate range of $\cSi$ for each bin (and divided by two to account for the banding structure in $\cSiib$):
\begin{equation}
  R_\mathrm{bin_i} = \frac{1}{2}\int_{\mathrm{lower}_i}^{\mathrm{upper}_i} \! \frac{\mathrm{d} R}{\mathrm{d}\cSi} \, \mathrm{d}\cSi
\end{equation}
With some simple rearrangement this rate can be written in terms of an integral over the detector response function $G$ as follows
\begin{align}
  R_\mathrm{bin_i} &= \frac{1}{2}\int\frac{\mathrm{d} R}{\mathrm{d}E}\int_{\mathrm{lower}_i}^{\mathrm{upper}_i} \! G(\cSi,E) \, \mathrm{d}\cSi \, \mathrm{d}E \\
 &= \int\frac{\mathrm{d} R}{\mathrm{d}E} G'_i(E) \mathrm{d}E
\end{align}
where in the last line we absorb the factor of $1/2$ into the definition of $G'_i$. We see here that the signal rate for each bin can be expressed as an integral over the recoil spectrum times a detector response function $G'_i$ for that bin. It is these detector response functions which are shown in Fig.~\ref{fig:smeartable_highE}, and which we provide digitally for use by the community. A low-resolution example is given in Table \ref{tab:smeartable_highE}. With these tables it is simple to produce a signal model for our analysis for any theoretical recoil spectrum. The functions $G'_i$ are provided for three values of the nuisance variable $\Leff$, namely the median value and values at $\pm 1 \sigma$ in $\Leff$. From these, along with the measured background rates given in table \ref{table:BinDef}, one may construct a likelihood which accounts for uncertainties in $\Leff$, Alternatively simply using the $-1\sigma$ value produces quite an accurate prediction and is generally conservative.

\begin{table}
{
  \lstset{tabsize=4,basicstyle=\tiny\ttfamily,columns=flexible,emptylines=10000,keepspaces=true}
\begin{lstlisting}
# E(keV) bin 1    bin 2    bin 3    bin 4    bin 5    bin 6    bin 7    bin 8    bin 9
3.00e+00 1.44e-22 2.70e-32 1.23e-42 0.00e+00 0.00e+00 0.00e+00 1.44e-22 1.23e-42 0.00e+00
1.30e+01 9.21e-09 7.58e-14 1.25e-19 6.21e-40 0.00e+00 0.00e+00 9.21e-09 1.25e-19 0.00e+00
2.30e+01 1.74e-04 1.07e-07 1.24e-11 1.51e-26 0.00e+00 0.00e+00 1.74e-04 1.24e-11 2.64e-32
3.30e+01 2.22e-02 2.79e-04 6.56e-07 5.47e-18 8.20e-38 0.00e+00 2.25e-02 6.56e-07 1.71e-22
4.30e+01 1.59e-01 1.68e-02 3.50e-04 1.89e-12 1.24e-28 1.82e-43 1.76e-01 3.50e-04 4.95e-16
5.30e+01 2.23e-01 1.21e-01 1.40e-02 1.28e-08 6.89e-22 1.43e-34 3.44e-01 1.40e-02 1.82e-11
6.30e+01 1.10e-01 2.12e-01 9.84e-02 4.73e-06 5.28e-17 5.47e-28 3.21e-01 9.84e-02 2.59e-08
7.30e+01 2.77e-02 1.54e-01 2.51e-01 2.58e-04 2.20e-13 5.56e-23 1.82e-01 2.51e-01 4.20e-06
8.30e+01 4.38e-03 6.14e-02 3.67e-01 4.07e-03 1.36e-10 5.26e-19 6.58e-02 3.71e-01 1.65e-04
9.30e+01 4.65e-04 1.52e-02 3.96e-01 2.73e-02 2.31e-08 1.01e-15 1.57e-02 4.21e-01 2.44e-03
1.03e+02 3.40e-05 2.47e-03 3.41e-01 9.81e-02 1.50e-06 6.05e-13 2.50e-03 4.21e-01 1.75e-02
1.13e+02 1.91e-06 2.89e-04 2.29e-01 2.13e-01 4.09e-05 1.22e-10 2.91e-04 3.74e-01 6.77e-02
1.23e+02 7.75e-08 2.38e-05 1.14e-01 3.28e-01 5.91e-04 1.16e-08 2.39e-05 2.76e-01 1.66e-01
1.33e+02 2.18e-09 1.33e-06 3.98e-02 3.97e-01 5.03e-03 5.94e-07 1.33e-06 1.55e-01 2.87e-01
1.43e+02 5.40e-11 6.21e-08 1.05e-02 4.06e-01 2.41e-02 1.42e-05 6.21e-08 6.64e-02 3.74e-01
1.53e+02 1.33e-12 2.71e-09 2.23e-03 3.66e-01 7.14e-02 1.73e-04 2.71e-09 2.26e-02 4.17e-01
1.63e+02 2.86e-14 1.00e-10 3.75e-04 2.85e-01 1.51e-01 1.32e-03 1.00e-10 6.04e-03 4.32e-01
1.73e+02 5.43e-16 3.19e-12 5.09e-05 1.86e-01 2.43e-01 6.76e-03 3.19e-12 1.28e-03 4.34e-01
1.83e+02 9.29e-18 8.90e-14 5.69e-06 1.01e-01 3.09e-01 2.42e-02 8.90e-14 2.21e-04 4.34e-01
1.93e+02 1.44e-19 2.21e-15 5.32e-07 4.46e-02 3.23e-01 6.38e-02 2.21e-15 3.14e-05 4.31e-01
2.03e+02 2.05e-21 4.92e-17 4.23e-08 1.62e-02 2.83e-01 1.29e-01 4.92e-17 3.73e-06 4.28e-01
2.13e+02 2.71e-23 9.96e-19 2.91e-09 4.89e-03 2.10e-01 2.06e-01 9.96e-19 3.78e-07 4.21e-01
2.23e+02 3.33e-25 1.85e-20 1.74e-10 1.23e-03 1.31e-01 2.71e-01 1.85e-20 3.29e-08 4.04e-01
2.33e+02 3.83e-27 3.16e-22 9.25e-12 2.63e-04 6.94e-02 2.99e-01 3.16e-22 2.51e-09 3.69e-01
2.43e+02 4.16e-29 5.03e-24 4.38e-13 4.80e-05 3.12e-02 2.81e-01 5.03e-24 1.68e-10 3.12e-01
2.53e+02 4.29e-31 7.48e-26 1.87e-14 7.55e-06 1.20e-02 2.27e-01 7.48e-26 1.00e-11 2.39e-01
2.63e+02 4.21e-33 1.05e-27 7.23e-16 1.04e-06 3.94e-03 1.58e-01 1.05e-27 5.38e-13 1.62e-01
2.73e+02 3.95e-35 1.39e-29 2.56e-17 1.25e-07 1.12e-03 9.59e-02 1.39e-29 2.61e-14 9.70e-02
2.83e+02 3.56e-37 1.74e-31 8.33e-19 1.34e-08 2.77e-04 5.04e-02 1.74e-31 1.15e-15 5.07e-02
2.93e+02 3.08e-39 2.08e-33 2.51e-20 1.29e-09 6.00e-05 2.31e-02 2.08e-33 4.67e-17 2.31e-02
3.03e+02 2.58e-41 2.38e-35 7.04e-22 1.11e-10 1.15e-05 9.25e-03 2.38e-35 1.75e-18 9.26e-03
3.13e+02 2.03e-43 2.61e-37 1.84e-23 8.69e-12 1.95e-06 3.26e-03 2.61e-37 6.06e-20 3.26e-03
3.23e+02 0.00e+00 2.76e-39 4.54e-25 6.20e-13 2.97e-07 1.01e-03 2.76e-39 1.96e-21 1.01e-03
3.33e+02 0.00e+00 2.81e-41 1.05e-26 4.06e-14 4.06e-08 2.80e-04 2.81e-41 5.93e-23 2.80e-04
3.43e+02 0.00e+00 2.72e-43 2.32e-28 2.44e-15 5.04e-09 6.91e-05 2.72e-43 1.69e-24 6.91e-05

\end{lstlisting}
}
\caption{Detector response table using $\Leff$ with constrained scaling parameter set to $-1\sigma$ value. First column gives recoil energies, subsequent columns give the values of $G'_i(E)$ for each of the 9 high-energy analysis bins. The sampling is in steps of $10~\keVr$, which is too coarse to give an accurate signal model for very low WIMP masses, but is suitable for the mass range most relevant to our analysis. Higher resolution $G'_i(E)$ functions, and $G'_i(E)$ functions for other values of $\Leff$, are given in supplementary material. 
\label{tab:smeartable_highE}
}
\end{table}  
\newpage
\subsection{Example code}
\label{app:example_code}
\begin{lstlisting}
import numpy as np
from numpy import newaxis
from scipy.interpolate import interp1d

def TrapI(x,y):
    """Simple trapezoid integration"""
    w = x[1:] - x[:-1]
    h = (y[1:] + y[:-1])/2.
    return np.sum(w*h,axis=0)

# Load detector response table
data = np.loadtxt("detector_table.dat")
E = data[:,0]; Gi = data[:,1:]
# Load test recoil spectrum (1 TeV WIMP, O6)
data = np.loadtxt("O6_1TeV.dat")
Er = data[:,0]
# Input spectra is normalised to coupling^2=1,
# rescale to something near limit (1e3)
# Also multiply in the appropriate exposure
dRdE = data[:,1] * (1e3/1.) * 224.6*34.
# Interpolate recoil spectrum to table values
# Assume spectrum zero outside data given
f_dRdE = interp1d(Er,dRdE)
dRdE_matched = f_dRdE(E)
Ri = TrapI(E[:,newaxis],Gi*dRdE_matched[:,newaxis])

for i,R in enumerate(Ri):
  print "bin {0}: rate = {1:.2g}".format(i+1,R)

Output:

bin 1: rate = 0.081
bin 2: rate = 0.098
bin 3: rate = 0.35
bin 4: rate = 0.46
bin 5: rate = 0.29
bin 6: rate = 0.22
bin 7: rate = 0.18
bin 8: rate = 0.47
bin 9: rate = 0.84
\end{lstlisting}

%\vfill

%%% BIBLIOGRAPHY %%%

\bibliography{EFTPaperBib}

\begin{thebibliography}{34}%
\makeatletter
\providecommand \@ifxundefined [1]{%
 \@ifx{#1\undefined}
}%
\providecommand \@ifnum [1]{%
 \ifnum #1\expandafter \@firstoftwo
 \else \expandafter \@secondoftwo
 \fi
}%
\providecommand \@ifx [1]{%
 \ifx #1\expandafter \@firstoftwo
 \else \expandafter \@secondoftwo
 \fi
}%
\providecommand \natexlab [1]{#1}%
\providecommand \enquote  [1]{``#1''}%
\providecommand \bibnamefont  [1]{#1}%
\providecommand \bibfnamefont [1]{#1}%
\providecommand \citenamefont [1]{#1}%
\providecommand \href@noop [0]{\@secondoftwo}%
\providecommand \href [0]{\begingroup \@sanitize@url \@href}%
\providecommand \@href[1]{\@@startlink{#1}\@@href}%
\providecommand \@@href[1]{\endgroup#1\@@endlink}%
\providecommand \@sanitize@url [0]{\catcode `\\12\catcode `\$12\catcode
  `\&12\catcode `\#12\catcode `\^12\catcode `\_12\catcode `\%12\relax}%
\providecommand \@@startlink[1]{}%
\providecommand \@@endlink[0]{}%
\providecommand \url  [0]{\begingroup\@sanitize@url \@url }%
\providecommand \@url [1]{\endgroup\@href {#1}{\urlprefix }}%
\providecommand \urlprefix  [0]{URL }%
\providecommand \Eprint [0]{\href }%
\providecommand \doibase [0]{http://dx.doi.org/}%
\providecommand \selectlanguage [0]{\@gobble}%
\providecommand \bibinfo  [0]{\@secondoftwo}%
\providecommand \bibfield  [0]{\@secondoftwo}%
\providecommand \translation [1]{[#1]}%
\providecommand \BibitemOpen [0]{}%
\providecommand \bibitemStop [0]{}%
\providecommand \bibitemNoStop [0]{.\EOS\space}%
\providecommand \EOS [0]{\spacefactor3000\relax}%
\providecommand \BibitemShut  [1]{\csname bibitem#1\endcsname}%
\let\auto@bib@innerbib\@empty
%</preamble>
\bibitem [{\citenamefont {Harvey}\ \emph {et~al.}(2015)\citenamefont {Harvey},
  \citenamefont {Massey}, \citenamefont {Kitching}, \citenamefont {Taylor},\
  and\ \citenamefont {Tittley}}]{Harvey1462}%
  \BibitemOpen
  \bibfield  {author} {\bibinfo {author} {\bibfnamefont {D.}~\bibnamefont
  {Harvey}}, \bibinfo {author} {\bibfnamefont {R.}~\bibnamefont {Massey}},
  \bibinfo {author} {\bibfnamefont {T.}~\bibnamefont {Kitching}}, \bibinfo
  {author} {\bibfnamefont {A.}~\bibnamefont {Taylor}}, \ and\ \bibinfo {author}
  {\bibfnamefont {E.}~\bibnamefont {Tittley}},\ }\href {\doibase
  10.1126/science.1261381} {\bibfield  {journal} {\bibinfo  {journal}
  {Science}\ }\textbf {\bibinfo {volume} {347}},\ \bibinfo {pages} {1462}
  (\bibinfo {year} {2015})},\ \Eprint {http://arxiv.org/abs/1503.07675}
  {arXiv:1503.07675 [astro-ph.CO]} \BibitemShut {NoStop}%
%%CITATION = ARXIV:1503.07675;%%
\bibitem [{\citenamefont {Bennett}\ \emph {et~al.}(2013)\citenamefont {Bennett}
  \emph {et~al.}}]{WMAP:9years}%
  \BibitemOpen
  \bibfield  {author} {\bibinfo {author} {\bibfnamefont {C.~L.}\ \bibnamefont
  {Bennett}} \emph {et~al.} (\bibinfo {collaboration} {WMAP}),\ }\href
  {\doibase 10.1088/0067-0049/208/2/20} {\bibfield  {journal} {\bibinfo
  {journal} {Astrophys. J. Suppl.}\ }\textbf {\bibinfo {volume} {208}},\
  \bibinfo {pages} {20} (\bibinfo {year} {2013})},\ \Eprint
  {http://arxiv.org/abs/1212.5225} {arXiv:1212.5225 [astro-ph.CO]} \BibitemShut
  {NoStop}%
%%CITATION = ARXIV:1212.5225;%%
\bibitem [{\citenamefont {Ade}\ \emph {et~al.}(2016)\citenamefont {Ade} \emph
  {et~al.}}]{PLANCK}%
  \BibitemOpen
  \bibfield  {author} {\bibinfo {author} {\bibfnamefont {P.~A.~R.}\
  \bibnamefont {Ade}} \emph {et~al.} (\bibinfo {collaboration} {Planck}),\
  }\href {\doibase 10.1051/0004-6361/201525830} {\bibfield  {journal} {\bibinfo
   {journal} {Astron. Astrophys.}\ }\textbf {\bibinfo {volume} {594}},\
  \bibinfo {pages} {A13} (\bibinfo {year} {2016})},\ \Eprint
  {http://arxiv.org/abs/1502.01589} {arXiv:1502.01589 [astro-ph.CO]}
  \BibitemShut {NoStop}%
%%CITATION = ARXIV:1502.01589;%%
\bibitem [{\citenamefont {Silk}\ \emph {et~al.}(2010)\citenamefont {Silk} \emph
  {et~al.}}]{Bertone:2010zza}%
  \BibitemOpen
  \bibfield  {author} {\bibinfo {author} {\bibfnamefont {J.}~\bibnamefont
  {Silk}} \emph {et~al.},\ }\href {\doibase 10.1017/CBO9780511770739} {\emph
  {\bibinfo {title} {{Particle Dark Matter: Observations, Models and
  Searches}}}}\ (\bibinfo  {publisher} {Cambridge Univ. Press},\ \bibinfo
  {address} {Cambridge},\ \bibinfo {year} {2010})\BibitemShut {NoStop}%
%%CITATION = INSPIRE-895273;%%
\bibitem [{\citenamefont {Aprile}\ \emph {et~al.}(2016)\citenamefont {Aprile}
  \emph {et~al.}}]{xe100_run_combination}%
  \BibitemOpen
  \bibfield  {author} {\bibinfo {author} {\bibfnamefont {E.}~\bibnamefont
  {Aprile}} \emph {et~al.} (\bibinfo {collaboration} {XENON100}),\ }\href
  {\doibase 10.1103/PhysRevD.94.122001} {\bibfield  {journal} {\bibinfo
  {journal} {Phys. Rev.}\ }\textbf {\bibinfo {volume} {D94}},\ \bibinfo {pages}
  {122001} (\bibinfo {year} {2016})},\ \Eprint
  {http://arxiv.org/abs/1609.06154} {arXiv:1609.06154 [astro-ph.CO]}
  \BibitemShut {NoStop}%
%%CITATION = ARXIV:1609.06154;%%
\bibitem [{\citenamefont {Tan}\ \emph {et~al.}(2016)\citenamefont {Tan} \emph
  {et~al.}}]{PANDAX}%
  \BibitemOpen
  \bibfield  {author} {\bibinfo {author} {\bibfnamefont {A.}~\bibnamefont
  {Tan}} \emph {et~al.} (\bibinfo {collaboration} {PandaX-II}),\ }\href
  {\doibase 10.1103/PhysRevLett.117.121303} {\bibfield  {journal} {\bibinfo
  {journal} {Phys. Rev. Lett.}\ }\textbf {\bibinfo {volume} {117}},\ \bibinfo
  {pages} {121303} (\bibinfo {year} {2016})},\ \Eprint
  {http://arxiv.org/abs/1607.07400} {arXiv:1607.07400 [hep-ex]} \BibitemShut
  {NoStop}%
%%CITATION = ARXIV:1607.07400;%%
\bibitem [{\citenamefont {Akerib}\ \emph {et~al.}(2016)\citenamefont {Akerib}
  \emph {et~al.}}]{LUXnew}%
  \BibitemOpen
  \bibfield  {author} {\bibinfo {author} {\bibfnamefont {D.~S.}\ \bibnamefont
  {Akerib}} \emph {et~al.} (\bibinfo {collaboration} {LUX}),\ }\href {\doibase
  10.1103/PhysRevLett.116.161301} {\bibfield  {journal} {\bibinfo  {journal}
  {Phys. Rev. Lett.}\ }\textbf {\bibinfo {volume} {116}},\ \bibinfo {pages}
  {161301} (\bibinfo {year} {2016})},\ \Eprint
  {http://arxiv.org/abs/1512.03506} {arXiv:1512.03506 [astro-ph.CO]}
  \BibitemShut {NoStop}%
%%CITATION = ARXIV:1512.03506;%%
\bibitem [{\citenamefont {Aalseth}\ \emph {et~al.}(2013)\citenamefont {Aalseth}
  \emph {et~al.}}]{COGENT}%
  \BibitemOpen
  \bibfield  {author} {\bibinfo {author} {\bibfnamefont {C.~E.}\ \bibnamefont
  {Aalseth}} \emph {et~al.} (\bibinfo {collaboration} {CoGeNT}),\ }\href
  {\doibase 10.1103/PhysRevD.88.012002} {\bibfield  {journal} {\bibinfo
  {journal} {Phys. Rev.}\ }\textbf {\bibinfo {volume} {D88}},\ \bibinfo {pages}
  {012002} (\bibinfo {year} {2013})},\ \Eprint {http://arxiv.org/abs/1208.5737}
  {arXiv:1208.5737 [astro-ph.CO]} \BibitemShut {NoStop}%
%%CITATION = ARXIV:1208.5737;%%
\bibitem [{\citenamefont {Agnese}\ \emph {et~al.}(2014)\citenamefont {Agnese}
  \emph {et~al.}}]{CDMSlite}%
  \BibitemOpen
  \bibfield  {author} {\bibinfo {author} {\bibfnamefont {R.}~\bibnamefont
  {Agnese}} \emph {et~al.} (\bibinfo {collaboration} {SuperCDMS
  collaboration}),\ }\href {\doibase 10.1103/PhysRevLett.112.041302} {\bibfield
   {journal} {\bibinfo  {journal} {Phys. Rev. Lett.}\ }\textbf {\bibinfo
  {volume} {112}},\ \bibinfo {pages} {041302} (\bibinfo {year}
  {2014})}\BibitemShut {NoStop}%
\bibitem [{\citenamefont {Angloher}\ \emph {et~al.}(2012)\citenamefont
  {Angloher} \emph {et~al.}}]{CREST}%
  \BibitemOpen
  \bibfield  {author} {\bibinfo {author} {\bibfnamefont {G.}~\bibnamefont
  {Angloher}} \emph {et~al.},\ }\href {\doibase 10.1140/epjc/s10052-012-1971-8}
  {\bibfield  {journal} {\bibinfo  {journal} {Eur. Phys. J.}\ }\textbf
  {\bibinfo {volume} {C72}},\ \bibinfo {pages} {1971} (\bibinfo {year}
  {2012})},\ \Eprint {http://arxiv.org/abs/1109.0702} {arXiv:1109.0702
  [astro-ph.CO]} \BibitemShut {NoStop}%
%%CITATION = ARXIV:1109.0702;%%
\bibitem [{\citenamefont {Bernabei}\ \emph {et~al.}(2010)\citenamefont
  {Bernabei} \emph {et~al.}}]{DAMA}%
  \BibitemOpen
  \bibfield  {author} {\bibinfo {author} {\bibfnamefont {R.}~\bibnamefont
  {Bernabei}} \emph {et~al.} (\bibinfo {collaboration} {DAMA, LIBRA}),\ }\href
  {\doibase 10.1140/epjc/s10052-010-1303-9} {\bibfield  {journal} {\bibinfo
  {journal} {Eur. Phys. J.}\ }\textbf {\bibinfo {volume} {C67}},\ \bibinfo
  {pages} {39} (\bibinfo {year} {2010})},\ \Eprint
  {http://arxiv.org/abs/1002.1028} {arXiv:1002.1028 [astro-ph.GA]} \BibitemShut
  {NoStop}%
%%CITATION = ARXIV:1002.1028;%%
\bibitem [{\citenamefont {Lewin}\ and\ \citenamefont {Smith}(1996)}]{LEWIN}%
  \BibitemOpen
  \bibfield  {author} {\bibinfo {author} {\bibfnamefont {J.~D.}\ \bibnamefont
  {Lewin}}\ and\ \bibinfo {author} {\bibfnamefont {P.~F.}\ \bibnamefont
  {Smith}},\ }\href {\doibase 10.1016/S0927-6505(96)00047-3} {\bibfield
  {journal} {\bibinfo  {journal} {Astropart. Phys.}\ }\textbf {\bibinfo
  {volume} {6}},\ \bibinfo {pages} {87} (\bibinfo {year} {1996})}\BibitemShut
  {NoStop}%
%%CITATION = APHYE,6,87;%%
\bibitem [{\citenamefont {Chang}\ \emph {et~al.}(2010)\citenamefont {Chang},
  \citenamefont {Pierce},\ and\ \citenamefont {Weiner}}]{Chang:2009yt}%
  \BibitemOpen
  \bibfield  {author} {\bibinfo {author} {\bibfnamefont {S.}~\bibnamefont
  {Chang}}, \bibinfo {author} {\bibfnamefont {A.}~\bibnamefont {Pierce}}, \
  and\ \bibinfo {author} {\bibfnamefont {N.}~\bibnamefont {Weiner}},\ }\href
  {\doibase 10.1088/1475-7516/2010/01/006} {\bibfield  {journal} {\bibinfo
  {journal} {JCAP}\ }\textbf {\bibinfo {volume} {1001}},\ \bibinfo {pages}
  {006} (\bibinfo {year} {2010})},\ \Eprint {http://arxiv.org/abs/0908.3192}
  {arXiv:0908.3192 [hep-ph]} \BibitemShut {NoStop}%
%%CITATION = ARXIV:0908.3192;%%
\bibitem [{\citenamefont {Fitzpatrick}\ \emph {et~al.}(2012)\citenamefont
  {Fitzpatrick}, \citenamefont {Haxton}, \citenamefont {Katz}, \citenamefont
  {Lubbers},\ and\ \citenamefont {Xu}}]{Fitzpatrick:2012ib}%
  \BibitemOpen
  \bibfield  {author} {\bibinfo {author} {\bibfnamefont {A.~L.}\ \bibnamefont
  {Fitzpatrick}}, \bibinfo {author} {\bibfnamefont {W.}~\bibnamefont {Haxton}},
  \bibinfo {author} {\bibfnamefont {E.}~\bibnamefont {Katz}}, \bibinfo {author}
  {\bibfnamefont {N.}~\bibnamefont {Lubbers}}, \ and\ \bibinfo {author}
  {\bibfnamefont {Y.}~\bibnamefont {Xu}},\ }\href@noop {} {\  (\bibinfo {year}
  {2012})},\ \Eprint {http://arxiv.org/abs/1211.2818} {arXiv:1211.2818
  [hep-ph]} \BibitemShut {NoStop}%
%%CITATION = ARXIV:1211.2818;%%
\bibitem [{\citenamefont {Anand}\ \emph {et~al.}(2014)\citenamefont {Anand},
  \citenamefont {Fitzpatrick},\ and\ \citenamefont {Haxton}}]{Anand:MathTools}%
  \BibitemOpen
  \bibfield  {author} {\bibinfo {author} {\bibfnamefont {N.}~\bibnamefont
  {Anand}}, \bibinfo {author} {\bibfnamefont {A.~L.}\ \bibnamefont
  {Fitzpatrick}}, \ and\ \bibinfo {author} {\bibfnamefont {W.~C.}\ \bibnamefont
  {Haxton}},\ }\href {\doibase 10.1103/PhysRevC.89.065501} {\bibfield
  {journal} {\bibinfo  {journal} {Phys. Rev.}\ }\textbf {\bibinfo {volume}
  {C89}},\ \bibinfo {pages} {065501} (\bibinfo {year} {2014})},\ \Eprint
  {http://arxiv.org/abs/1308.6288} {arXiv:1308.6288 [hep-ph]} \BibitemShut
  {NoStop}%
%%CITATION = ARXIV:1308.6288;%%
\bibitem [{\citenamefont {Fitzpatrick}\ \emph {et~al.}(2013)\citenamefont
  {Fitzpatrick}, \citenamefont {Haxton}, \citenamefont {Katz}, \citenamefont
  {Lubbers},\ and\ \citenamefont {Xu}}]{Fitzpatrick:MathTools}%
  \BibitemOpen
  \bibfield  {author} {\bibinfo {author} {\bibfnamefont {A.~L.}\ \bibnamefont
  {Fitzpatrick}}, \bibinfo {author} {\bibfnamefont {W.}~\bibnamefont {Haxton}},
  \bibinfo {author} {\bibfnamefont {E.}~\bibnamefont {Katz}}, \bibinfo {author}
  {\bibfnamefont {N.}~\bibnamefont {Lubbers}}, \ and\ \bibinfo {author}
  {\bibfnamefont {Y.}~\bibnamefont {Xu}},\ }\href {\doibase
  10.1088/1475-7516/2013/02/004} {\bibfield  {journal} {\bibinfo  {journal}
  {JCAP}\ }\textbf {\bibinfo {volume} {1302}},\ \bibinfo {pages} {004}
  (\bibinfo {year} {2013})},\ \Eprint {http://arxiv.org/abs/1203.3542}
  {arXiv:1203.3542 [hep-ph]} \BibitemShut {NoStop}%
%%CITATION = ARXIV:1203.3542;%%
\bibitem [{\citenamefont {Tucker-Smith}\ and\ \citenamefont
  {Weiner}(2001)}]{InelasticIntro}%
  \BibitemOpen
  \bibfield  {author} {\bibinfo {author} {\bibfnamefont {D.}~\bibnamefont
  {Tucker-Smith}}\ and\ \bibinfo {author} {\bibfnamefont {N.}~\bibnamefont
  {Weiner}},\ }\href {\doibase 10.1103/PhysRevD.64.043502} {\bibfield
  {journal} {\bibinfo  {journal} {Phys. Rev.}\ }\textbf {\bibinfo {volume}
  {D64}},\ \bibinfo {pages} {043502} (\bibinfo {year} {2001})},\ \Eprint
  {http://arxiv.org/abs/hep-ph/0101138} {arXiv:hep-ph/0101138 [hep-ph]}
  \BibitemShut {NoStop}%
%%CITATION = HEP-PH/0101138;%%
\bibitem [{\citenamefont {Barello}\ \emph {et~al.}(2014)\citenamefont
  {Barello}, \citenamefont {Chang},\ and\ \citenamefont
  {Newby}}]{InelasticMath}%
  \BibitemOpen
  \bibfield  {author} {\bibinfo {author} {\bibfnamefont {G.}~\bibnamefont
  {Barello}}, \bibinfo {author} {\bibfnamefont {S.}~\bibnamefont {Chang}}, \
  and\ \bibinfo {author} {\bibfnamefont {C.~A.}\ \bibnamefont {Newby}},\ }\href
  {\doibase 10.1103/PhysRevD.90.094027} {\bibfield  {journal} {\bibinfo
  {journal} {Phys. Rev.}\ }\textbf {\bibinfo {volume} {D90}},\ \bibinfo {pages}
  {094027} (\bibinfo {year} {2014})},\ \Eprint {http://arxiv.org/abs/1409.0536}
  {arXiv:1409.0536 [hep-ph]} \BibitemShut {NoStop}%
%%CITATION = ARXIV:1409.0536;%%
\bibitem [{\citenamefont {Aprile}\ \emph
  {et~al.}(2012{\natexlab{a}})\citenamefont {Aprile} \emph
  {et~al.}}]{xe100_instr2012}%
  \BibitemOpen
  \bibfield  {author} {\bibinfo {author} {\bibfnamefont {E.}~\bibnamefont
  {Aprile}} \emph {et~al.} (\bibinfo {collaboration} {XENON100}),\ }\href
  {\doibase 10.1016/j.astropartphys.2012.01.003} {\bibfield  {journal}
  {\bibinfo  {journal} {Astropart. Phys.}\ }\textbf {\bibinfo {volume} {35}},\
  \bibinfo {pages} {573} (\bibinfo {year} {2012}{\natexlab{a}})},\ \Eprint
  {http://arxiv.org/abs/1107.2155} {arXiv:1107.2155 [astro-ph.IM]} \BibitemShut
  {NoStop}%
%%CITATION = ARXIV:1107.2155;%%
\bibitem [{\citenamefont {Aprile}\ \emph
  {et~al.}(2012{\natexlab{b}})\citenamefont {Aprile} \emph
  {et~al.}}]{xe100_run10_si}%
  \BibitemOpen
  \bibfield  {author} {\bibinfo {author} {\bibfnamefont {E.}~\bibnamefont
  {Aprile}} \emph {et~al.} (\bibinfo {collaboration} {XENON100}),\ }\href
  {\doibase 10.1103/PhysRevLett.109.181301} {\bibfield  {journal} {\bibinfo
  {journal} {Phys. Rev. Lett.}\ }\textbf {\bibinfo {volume} {109}},\ \bibinfo
  {pages} {181301} (\bibinfo {year} {2012}{\natexlab{b}})},\ \Eprint
  {http://arxiv.org/abs/1207.5988} {arXiv:1207.5988 [astro-ph.CO]} \BibitemShut
  {NoStop}%
%%CITATION = ARXIV:1207.5988;%%
\bibitem [{\citenamefont {Aprile}\ \emph
  {et~al.}(2014{\natexlab{a}})\citenamefont {Aprile} \emph
  {et~al.}}]{Aprile:2012vw}%
  \BibitemOpen
  \bibfield  {author} {\bibinfo {author} {\bibfnamefont {E.}~\bibnamefont
  {Aprile}} \emph {et~al.} (\bibinfo {collaboration} {XENON100}),\ }\href
  {\doibase 10.1016/j.astropartphys.2013.10.002} {\bibfield  {journal}
  {\bibinfo  {journal} {Astropart. Phys.}\ }\textbf {\bibinfo {volume} {54}},\
  \bibinfo {pages} {11} (\bibinfo {year} {2014}{\natexlab{a}})},\ \Eprint
  {http://arxiv.org/abs/1207.3458} {arXiv:1207.3458 [astro-ph.IM]} \BibitemShut
  {NoStop}%
%%CITATION = ARXIV:1207.3458;%%
\bibitem [{\citenamefont {Aprile}\ \emph
  {et~al.}(2013{\natexlab{a}})\citenamefont {Aprile} \emph
  {et~al.}}]{Aprile:2013tov}%
  \BibitemOpen
  \bibfield  {author} {\bibinfo {author} {\bibfnamefont {E.}~\bibnamefont
  {Aprile}} \emph {et~al.} (\bibinfo {collaboration} {XENON100}),\ }\href
  {\doibase 10.1088/0954-3899/40/11/115201} {\bibfield  {journal} {\bibinfo
  {journal} {J. Phys.}\ }\textbf {\bibinfo {volume} {G40}},\ \bibinfo {pages}
  {115201} (\bibinfo {year} {2013}{\natexlab{a}})},\ \Eprint
  {http://arxiv.org/abs/1306.2303} {arXiv:1306.2303 [astro-ph.IM]} \BibitemShut
  {NoStop}%
%%CITATION = ARXIV:1306.2303;%%
\bibitem [{\citenamefont {Aprile}\ \emph
  {et~al.}(2013{\natexlab{b}})\citenamefont {Aprile} \emph
  {et~al.}}]{DataMCXenon}%
  \BibitemOpen
  \bibfield  {author} {\bibinfo {author} {\bibfnamefont {E.}~\bibnamefont
  {Aprile}} \emph {et~al.} (\bibinfo {collaboration} {XENON100}),\ }\href
  {\doibase 10.1103/PhysRevD.88.012006} {\bibfield  {journal} {\bibinfo
  {journal} {Phys. Rev.}\ }\textbf {\bibinfo {volume} {D88}},\ \bibinfo {pages}
  {012006} (\bibinfo {year} {2013}{\natexlab{b}})},\ \Eprint
  {http://arxiv.org/abs/1304.1427} {arXiv:1304.1427 [astro-ph.IM]} \BibitemShut
  {NoStop}%
%%CITATION = ARXIV:1304.1427;%%
\bibitem [{\citenamefont {Aprile}\ \emph
  {et~al.}(2014{\natexlab{b}})\citenamefont {Aprile} \emph
  {et~al.}}]{XenonSingleElectron}%
  \BibitemOpen
  \bibfield  {author} {\bibinfo {author} {\bibfnamefont {E.}~\bibnamefont
  {Aprile}} \emph {et~al.} (\bibinfo {collaboration} {XENON100}),\ }\href
  {\doibase 10.1088/0954-3899/41/3/035201} {\bibfield  {journal} {\bibinfo
  {journal} {J. Phys.}\ }\textbf {\bibinfo {volume} {G41}},\ \bibinfo {pages}
  {035201} (\bibinfo {year} {2014}{\natexlab{b}})},\ \Eprint
  {http://arxiv.org/abs/1311.1088} {arXiv:1311.1088 [physics.ins-det]}
  \BibitemShut {NoStop}%
%%CITATION = ARXIV:1311.1088;%%
\bibitem [{\citenamefont {Aprile}\ \emph
  {et~al.}(2013{\natexlab{c}})\citenamefont {Aprile} \emph
  {et~al.}}]{xe100_run10_sd}%
  \BibitemOpen
  \bibfield  {author} {\bibinfo {author} {\bibfnamefont {E.}~\bibnamefont
  {Aprile}} \emph {et~al.} (\bibinfo {collaboration} {XENON100
  collaboration}),\ }\href@noop {} {\bibfield  {journal} {\bibinfo  {journal}
  {Physical review letters}\ }\textbf {\bibinfo {volume} {111}},\ \bibinfo
  {pages} {021301} (\bibinfo {year} {2013}{\natexlab{c}})}\BibitemShut
  {NoStop}%
\bibitem [{\citenamefont {Cowan}\ \emph {et~al.}(2011)\citenamefont {Cowan},
  \citenamefont {Cranmer}, \citenamefont {Gross},\ and\ \citenamefont
  {Vitells}}]{asympt}%
  \BibitemOpen
  \bibfield  {author} {\bibinfo {author} {\bibfnamefont {G.}~\bibnamefont
  {Cowan}}, \bibinfo {author} {\bibfnamefont {K.}~\bibnamefont {Cranmer}},
  \bibinfo {author} {\bibfnamefont {E.}~\bibnamefont {Gross}}, \ and\ \bibinfo
  {author} {\bibfnamefont {O.}~\bibnamefont {Vitells}},\ }\href {\doibase
  10.1140/epjc/s10052-011-1554-0, 10.1140/epjc/s10052-013-2501-z} {\bibfield
  {journal} {\bibinfo  {journal} {Eur. Phys. J.}\ }\textbf {\bibinfo {volume}
  {C71}},\ \bibinfo {pages} {1554} (\bibinfo {year} {2011})},\ \bibinfo {note}
  {[Erratum: Eur. Phys. J.C73,2501(2013)]},\ \Eprint
  {http://arxiv.org/abs/1007.1727} {arXiv:1007.1727 [physics.data-an]}
  \BibitemShut {NoStop}%
%%CITATION = ARXIV:1007.1727;%%
\bibitem [{\citenamefont {Aprile}\ \emph
  {et~al.}(2011{\natexlab{a}})\citenamefont {Aprile} \emph
  {et~al.}}]{run8Result}%
  \BibitemOpen
  \bibfield  {author} {\bibinfo {author} {\bibfnamefont {E.}~\bibnamefont
  {Aprile}} \emph {et~al.} (\bibinfo {collaboration} {XENON100}),\ }\href
  {\doibase 10.1103/PhysRevLett.107.131302} {\bibfield  {journal} {\bibinfo
  {journal} {Phys. Rev. Lett.}\ }\textbf {\bibinfo {volume} {107}},\ \bibinfo
  {pages} {131302} (\bibinfo {year} {2011}{\natexlab{a}})},\ \Eprint
  {http://arxiv.org/abs/1104.2549} {arXiv:1104.2549 [astro-ph.CO]} \BibitemShut
  {NoStop}%
%%CITATION = ARXIV:1104.2549;%%
\bibitem [{\citenamefont {Aprile}\ \emph
  {et~al.}(2011{\natexlab{b}})\citenamefont {Aprile} \emph
  {et~al.}}]{Aprile:2011hx}%
  \BibitemOpen
  \bibfield  {author} {\bibinfo {author} {\bibfnamefont {E.}~\bibnamefont
  {Aprile}} \emph {et~al.} (\bibinfo {collaboration} {XENON100}),\ }\href
  {\doibase 10.1103/PhysRevD.84.052003} {\bibfield  {journal} {\bibinfo
  {journal} {Phys. Rev.}\ }\textbf {\bibinfo {volume} {D84}},\ \bibinfo {pages}
  {052003} (\bibinfo {year} {2011}{\natexlab{b}})},\ \Eprint
  {http://arxiv.org/abs/1103.0303} {arXiv:1103.0303 [hep-ex]} \BibitemShut
  {NoStop}%
%%CITATION = ARXIV:1103.0303;%%
\bibitem [{\citenamefont {Read}(2000)}]{cls}%
  \BibitemOpen
  \bibfield  {author} {\bibinfo {author} {\bibfnamefont {A.~L.}\ \bibnamefont
  {Read}},\ }in\ \href {http://weblib.cern.ch/abstract?CERN-OPEN-2000-205}
  {\emph {\bibinfo {booktitle} {{Workshop on confidence limits, CERN, Geneva,
  Switzerland, 17-18 Jan 2000: Proceedings}}}}\ (\bibinfo {year} {2000})\ pp.\
  \bibinfo {pages} {81--101}\BibitemShut {NoStop}%
%%CITATION = CERN-OPEN-2000-205;%%
\bibitem [{\citenamefont {Schneck}\ \emph {et~al.}(2015)\citenamefont {Schneck}
  \emph {et~al.}}]{CDMSEFT}%
  \BibitemOpen
  \bibfield  {author} {\bibinfo {author} {\bibfnamefont {K.}~\bibnamefont
  {Schneck}} \emph {et~al.} (\bibinfo {collaboration} {SuperCDMS}),\ }\href
  {\doibase 10.1103/PhysRevD.91.092004} {\bibfield  {journal} {\bibinfo
  {journal} {Phys. Rev.}\ }\textbf {\bibinfo {volume} {D91}},\ \bibinfo {pages}
  {092004} (\bibinfo {year} {2015})},\ \Eprint
  {http://arxiv.org/abs/1503.03379} {arXiv:1503.03379 [astro-ph.CO]}
  \BibitemShut {NoStop}%
%%CITATION = ARXIV:1503.03379;%%
\bibitem [{\citenamefont {Ahmed}\ \emph {et~al.}(2010)\citenamefont {Ahmed}
  \emph {et~al.}}]{CDMS_Inelastic}%
  \BibitemOpen
  \bibfield  {author} {\bibinfo {author} {\bibfnamefont {Z.}~\bibnamefont
  {Ahmed}} \emph {et~al.} (\bibinfo {collaboration} {CDMS-II}),\ }\href
  {\doibase 10.1126/science.1186112} {\bibfield  {journal} {\bibinfo  {journal}
  {Science}\ }\textbf {\bibinfo {volume} {327}},\ \bibinfo {pages} {1619}
  (\bibinfo {year} {2010})},\ \Eprint {http://arxiv.org/abs/0912.3592}
  {arXiv:0912.3592 [astro-ph.CO]} \BibitemShut {NoStop}%
%%CITATION = ARXIV:0912.3592;%%
\bibitem [{\citenamefont {Akimov}\ \emph {et~al.}(2010)\citenamefont {Akimov}
  \emph {et~al.}}]{Zepplin_Inel}%
  \BibitemOpen
  \bibfield  {author} {\bibinfo {author} {\bibfnamefont {D.~{\relax Yu}.}\
  \bibnamefont {Akimov}} \emph {et~al.} (\bibinfo {collaboration}
  {ZEPLIN-III}),\ }\href {\doibase 10.1016/j.physletb.2010.07.042} {\bibfield
  {journal} {\bibinfo  {journal} {Phys. Lett.}\ }\textbf {\bibinfo {volume}
  {B692}},\ \bibinfo {pages} {180} (\bibinfo {year} {2010})},\ \Eprint
  {http://arxiv.org/abs/1003.5626} {arXiv:1003.5626 [hep-ex]} \BibitemShut
  {NoStop}%
%%CITATION = ARXIV:1003.5626;%%
\bibitem [{\citenamefont {Aprile}\ \emph
  {et~al.}(2011{\natexlab{c}})\citenamefont {Aprile} \emph
  {et~al.}}]{XENON_Inelastic_WIMP}%
  \BibitemOpen
  \bibfield  {author} {\bibinfo {author} {\bibfnamefont {E.}~\bibnamefont
  {Aprile}} \emph {et~al.} (\bibinfo {collaboration} {XENON100}),\ }\href
  {\doibase 10.1103/PhysRevD.84.061101} {\bibfield  {journal} {\bibinfo
  {journal} {Phys. Rev.}\ }\textbf {\bibinfo {volume} {D84}},\ \bibinfo {pages}
  {061101} (\bibinfo {year} {2011}{\natexlab{c}})},\ \Eprint
  {http://arxiv.org/abs/1104.3121} {arXiv:1104.3121 [astro-ph.CO]} \BibitemShut
  {NoStop}%
%%CITATION = ARXIV:1104.3121;%%
\bibitem [{\citenamefont {De~Simone}\ and\ \citenamefont
  {Jacques}(2016)}]{DeSimone:2016fbz}%
  \BibitemOpen
  \bibfield  {author} {\bibinfo {author} {\bibfnamefont {A.}~\bibnamefont
  {De~Simone}}\ and\ \bibinfo {author} {\bibfnamefont {T.}~\bibnamefont
  {Jacques}},\ }\href {\doibase 10.1140/epjc/s10052-016-4208-4} {\bibfield
  {journal} {\bibinfo  {journal} {Eur. Phys. J.}\ }\textbf {\bibinfo {volume}
  {C76}},\ \bibinfo {pages} {367} (\bibinfo {year} {2016})},\ \Eprint
  {http://arxiv.org/abs/1603.08002} {arXiv:1603.08002 [hep-ph]} \BibitemShut
  {NoStop}%
%%CITATION = ARXIV:1603.08002;%%
\end{thebibliography}%

\end{document}